\documentclass[usenatbib, useAMS]{mn2e}

\usepackage{amsmath}
\usepackage[pdftex]{graphicx}
\usepackage{amssymb}
\usepackage[authoryear]{natbib}
\usepackage{times}

\makeatletter

\providecommand{\tabularnewline}{\\}

\newcommand{\aap}{A\&A}
\newcommand{\aaps}{A\&AS}
\newcommand{\apj}{ApJ}
\newcommand{\apjl}{\apj}

\newcommand{\aj}{AJ}
\newcommand{\mnras}{MNRAS}
\newcommand{\apjs}{ApJS}

\newcommand{\pasj}{PASJ}

\newcommand{\kmps}{\mathrm{km~s^{-1}}}
\newcommand{\ion}[2]{#1$\,${\sc {#2}}}   
\newcommand{\Kelvin}{\mathrm{K}}
\newcommand{\Msun}{\mathrm{M_{\sun}}}
\newcommand{\Rsun}{\mathrm{R_{\sun}}}

\newcommand{\MsunPerYear}{\mathrm{M_{\sun}\,yr^{-1}}}
\newcommand{\gammaHeI}{\gamma_{\mathrm{He\,\textsc{i}}}}

\font\tenbg=cmmib10 at 10pt
\def \rvecmu{{\hbox{\tenbg\char'026}}}

\newbox\grsign \setbox\grsign=\hbox{$>$} \newdimen\grdimen
\grdimen=\ht\grsign
\newbox\simlessbox \newbox\simgreatbox
\setbox\simgreatbox=\hbox{\raise.5ex\hbox{$>$}\llap
     {\lower.5ex\hbox{$\sim$}}}\ht1=\grdimen\dp1=0pt
\setbox\simlessbox=\hbox{\raise.5ex\hbox{$<$}\llap
     {\lower.5ex\hbox{$\sim$}}}\ht2=\grdimen\dp2=0pt
\def\simgreat{\mathrel{\copy\simgreatbox}}
\def\simless{\mathrel{\copy\simlessbox}}

\makeatother

\begin{document}
\title[Line profile models of the inner wind of CTTSs]{Line formation
  in the inner winds of classical T Tauri stars: testing the
  conical-shell wind solution}  

\author[Kurosawa \& Romanova]{Ryuichi
  Kurosawa\thanks{E-mail:kurosawa@astro.cornell.edu} and M.~M.~Romanova
  \\ Department of Astronomy, Cornell University, Space Sciences
  Building, Ithaca, NY 14853-6801, USA} 

\date{Dates to be inserted}

\maketitle

\begin{abstract}

We present the emission line profile models of hydrogen and helium
based on the results from axisymmetric magnetohydrodynamics (MHD)
simulations of the wind formed near the disk-magnetosphere boundary of
classical T Tauri stars (CTTSs). We extend the previous outflow models
of `the conical-shell wind' by Romanova et al.\,to include a well
defined magnetospheric accretion funnel flow which is essential for
modelling the optical and near-infrared hydrogen and helium lines of
CTTSs. The MHD model with an intermediate mass-accretion rate shows
outflows in conical-shell shape with a half opening angle $\sim
35^\circ$. The flow properties such as the maximum outflow speed in
the conical-shell wind, maximum inflow speed in the accretion funnel,
mass-accretion and mass-loss rates are comparable to those found in a
typical CTTS.  The density, velocity and modified
temperature from the MHD simulations are used in a separate radiative
transfer model to predict the line profiles and test the consistency
of the MHD models with observations. The line profiles are computed
with various combinations of X-ray luminosities, temperatures of X-ray
emitting plasma, and inclination angles. A rich diversity of line
profile morphology is found, and many of the model profiles are very
similar to those found in observations.  
We find that the conical-shell wind may contribute to the emission in
some hydrogen lines (e.g. H$\alpha$, H$\beta$, Pa$\beta$ and
Pa$\gamma$) significantly when the temperature in the wind is
relatively high (e.g. $\sim 10^{4}\,\Kelvin$); however, the wind
contribution decreases rapidly when a lower wind temperature is
adopted. The model well reproduces a relatively narrow and low-velocity
blueshifted absorption component in \ion{He}{i}~$\lambda$10830, which
are often seen in observations.

\end{abstract}


\section{Introduction}

\label{sec:intro}

Mass outflows are commonly found during mass accreting phases of
pre-main-sequence stars. While powerful collimated outflows are
observed in protostars, often less collimated and weaker outflows are
associated with classical T~Tauri stars (CTTSs) (e.g.~\citealt{Ray:2007}).  The
outflows in CTTSs are most likely powered by accretion processes, as the
strengths of mass outflow indicators are known to correlate with those
of mass accretion (e.g.~\citealt{cabrit:1990};
\citealt*{hartigan:1995}).  Understanding the origin and properties of
the wind from young stellar objects is very important as it is closely related
to their spin and disc evolutions. The mass-loss
process is one of the key mechanisms through which YSOs can remove
their angular momentum to the surrounding environment during the
active accretion phases (e.g.~\citealt{hartmann:1989};
\citealt{koenigl:1991}; \citealt{shu:1994}; \citealt{matt:2005};
\citeyear{Matt:2007b}; \citeyear{Matt:2008a}).

The outflows in the low-mass stars such as CTTSs are unlikely driven
by thermally or radiatively due to their relatively low temperature
and luminosity.  On other hand, relatively large magnetic field
strengths ($\sim 10^{3}$G) are found in CTTSs
(e.g.~\citealt{johns-krull:1999}; \citealt{symington:2005b};
\citealt{donati:2007}; \citealt{donati:2008}) which supports an idea
that the wind is formed in magnetohydrodynamics processes
(\citealt*{Koenigl:2011}).  The recent 
magnetohydrodynamics (MHD) simulation of \citet{Romanova:2009}
(hereafter RO09) showed a new type of outflow solution, so-called `the
conical-shell wind', which is produced when the stellar dipole magnetic field is compressed
by the accretion disc into the X-wind (\citealt{shu:1994}) like
configuration. The resulting outflow occurs in rather narrow conical-shell
shapes with their half opening angles between $30^{\circ}$ to
$40^{\circ}$. This conical-shell wind is similar to the X-wind of
\citet{shu:1994} in which the wind launching region is restricted to a
narrow region near the inner edge of the accretion disc around the
corotation radius.  However, unlike the X-wind model, the conical-shell wind
model of RO09 does not require the magnetospheric radius to be matched
with the corotation radius of the system.  More recent simulations
performed with larger radial domains have shown that the conical-shell winds
become strongly collimated by the magnetic force at larger distances
(\citealt*{Lii:2012}). The model has been also applied to the higher
mass-accretion rate FU~Orionis systems (FUORs) \citet{Koenigl:2011}.
Other possible wind configurations often considered for the outflows
in CTTSs (e.g.~\citealt*{Ferreira:2006}) are: 
(1)~a disc wind,  which requires a sufficient magnetic field and ionization
fraction to launch magnetocentrifugal winds over a relatively wide
range of disk radii (e.g.~\citealt{ustyugova:1995}; \citealt{Romanova:1997};
\citealt{ouyed:1997}; \citealt{Ustyugova:1999};
\citealt{koenigl:2000}; \citealt*{krasnopolsky:2003};
\citealt{Pudritz:2007}),  and (3)~a stellar
wind in which outflows occurs along the open magnetic fields from the
stellar surface (e.g., \citealt*{hartmann:1982}; \citealt{Kwan:1988};
\citealt{Hirose:1997}; \citealt{Romanova:2005};
\citealt{Cranmer:2009}).

Spectroscopic studies of strong emission lines in CTTSs may be able to
constrain the launching regions of the winds. This, in turn, would
restrict possible wind formation theories mentioned above.  A
usefulness of the optically thick \ion{He}{i}~$\lambda10830$ as a
diagnostic of the inner wind from the accreting stars has been
demonstrated in the observations by e.g.~\citet{Takami:2002},
\citet{Edwards:2003}, \citet{dupree:2005}, \citet{Edwards:2006}
(hereafter ED06) and \citet{Podio:2008}.  In particular,
ED06 showed that a very high fraction ($\sim 70$~per~cent) of CTTSs
exhibit a blueshifted absorption (below the continuum), which is clear sign
of outflow. This is much higher than the rate found in another wind
sensitive line, H$\alpha$, for which only about 10~per~cent of stars
exhibit a similar type of blueshifted absorption component (below the
continuum) (e.g., \citealt*{reipurth:1996}). \citet{Edwards:2007} and
\citet*{Kwan:2007} suggested that the blueshifted absorption component
in the \ion{He}{i}~$\lambda10830$ profiles is cased by a stellar wind
in about 40~per~cent, and by a disc wind in about 30~per~cent of the
samples in ED06. Using local excitation calculations combined with
spectroscopic observations, \citet{Kwan:2011} also  demonstrated the
effectiveness of \ion{He}{i}~$\lambda10830$ for probing the density
and temperature of the inner wind. The signs for the inner winds are
also seen blueshifted absorption component in Na~D, Ca~II~(H and K), and
Mg~II ($h$ and $k$) (e.g.~\citealt{Calvet:1997};
\citealt{Ardila:2002}); however, probing the launching regions
using these lines are more difficult (ED06).

Previously, we have developed and tested a multidimensional non-LTE radiative
transfer model of hydrogen and helium line profiles, which uses the
Sobolev approximation in the source function calculations
(\citealt*{Kurosawa:2011}, hereafter KU11; see also
\citealt{harries:2000}; \citealt*{kurosawa:2006}). 
The model uses rather simple
kinematic and geometric models for both stellar and disc winds (see
also \citealt{Hartmann:1990}; \citealt*{calvet:1992b}). We
demonstrated that our models are consistent with the scenario in which
the narrow blueshifted absorption component of
\ion{He}{i}~$\lambda10830$ seen in observations is caused by a disc
wind, and the wider blueshifted absorption component (the P-Cygni
profile) is caused by a bipolar stellar wind. Similar conclusions were
reached in the earlier study by \citet{Kwan:2007} who also applied
simple kinematic and geometric wind models. To advance our
understanding of the line formations in the inner wind of CTTSs, we now
apply our model to more realistic wind configurations such as those
found in MHD simulations. Therefore, our main goals in this paper are
to present the optical and near-infrared hydrogen and helium line
profile models computed with the density, velocity and
modified temperature (Section~\ref{sub:depend-Tmax}) from
the conical-shell wind solution found in the MHD simulations of RO09, and to
check the consistency of the wind solution with observed line profiles.

In Section~\ref{sec:MHD-model}, the model assumption and setup of our
MHD simulations are presented, and those for the radiative transfer
model for computing line profiles are given in
Section~\ref{sec:RT-model}.  The results of the MHD simulations and
the line profiles models are summarised in
Sections~\ref{sec:mhd-results} and \ref{sec:results-profiles},
respectively.  Brief discussions in which the line profile models are
compared with observations are given in Section~\ref{sec:discussion}. 
Finally main findings and conclusions are summarised in
Section~\ref{sec:conclusions}.

\section{MHD Model}

\label{sec:MHD-model}

The numerical method and the code used to simulate the outflow from
the disk-magnetosphere interaction region are essentially identical to
those in the conical-shell wind models of \citet{Romanova:2009} (hereafter
RO09; see also \citealt{Ustyugova:2006}); hence, we briefly describe
only the most important aspects of our numerical MHD simulations
below.

\subsection{Outline}

\label{sub:mhd-equations}

The simulation domain is divided into two regions: (1)~the accretion
disc region and (2)~the corona region which are located outside of the
disc region. The former is described by the equations of viscous and
resistive MHD, and the latter by those of ideal MHD. The plasma is
assumed to be an ideal gas with adiabatic index $\gamma=5/3$.  We
adopt the standard $\alpha$ prescriptions of \citet{shakura:1973} for
the coefficient of the turbulent kinematic viscosity
$\nu_{\mathrm{t}}\equiv\alpha_{\mathrm{v}}c_{\mathrm{s}}^{2}/\Omega_{\mathrm{K}}$,
where $c_{\mathrm{s}}$, $\Omega_{\mathrm{K}}$, $\alpha_{v}$ are the
isothermal sound speed, the Keplerian angular velocity, and the
dimensionless viscosity coefficient, respectively. Similarly, the
turbulent magnetic diffusivity coefficient is defined as
$\eta_{\mathrm{t}}\equiv\alpha_{\mathrm{d}}c_{\mathrm{s}}^{2}/\Omega_{\mathrm{K}}$
where $\alpha_{\mathrm{d}}$ are the dimensionless turbulent
diffusivity coefficient.  Both, viscosity and diffusivity correspond
to some type of turbulence in the disc; hence, they are included only
in the disc which is distinguished from the low-density corona. Thus,
we set $\alpha_\mathrm{v}, \alpha_{\rm d}=0$ when the density is lower
than a typical density in the disc ($\rho_\mathrm{d}$), i.e.~when
$\rho<\rho_\mathrm{d}$. The viscosity term helps to bring matter
towards the star in a steady rate, while diffusivity helps this matter
to diffuse through the magnetic field lines of the magnetosphere. We
refer to RO09 and \citet{Lii:2012} for more comprehensive descriptions
of our implementations of the stress tensor, viscosity and
diffusivity.

The MHD equations are solved numerically by using the
Godunov-type conservative code described in detail in RO09 and
\citet{Ustyugova:2006}. The code uses the spherical coordinates
$(r,\theta,\phi)$ where $\theta$ and $\phi$ are the polar and
azimuthal angles. We consider axisymmetric approximation,
i.e.~$\partial/\partial\phi=0$.

\subsection{Initial and boundary conditions}

\label{sub:init-bound-conditions}

The initial configuration of the magnetic field is set by a dipole
field aligned with the stellar spin axis, i.e. 

\begin{equation}
 \boldsymbol{B}=[3(\rvecmu\cdot\boldsymbol{r})\boldsymbol{r}-\rvecmu\,{r}^{2}]/r^{5},
 \label{eq:B-dipole}
\end{equation}
where $\rvecmu$ is the magnetic dipole moment of the star. The
simulation domain is initialised with a low-density, high-temperature
plasma (corona) which is described as the following density and
pressure distributions: $\rho=\rho_{\mathrm{c}}\exp[GM_{*}/({\cal
    R}T_{\mathrm{c}}r)]$ and $p=p_{\mathrm{c}}\exp[GM_{*}/({\cal
    R}T_{\mathrm{c}}r)]$ where $T_{\mathrm{c}}$, $\rho_{\mathrm{c}}$
and $p_{\mathrm{c}}$ are the temperature, density and pressure at the
outer boundary, respectively. The symbols $G$ and $M_{*}$ are the
gravitational constant and the stellar mass. The initial velocity of
the corona is set to zero everywhere.  The outer boundary
$r=R_{\mathrm{out}}$ are divided into a disc region
$\theta_{\mathrm{d}}\leq\theta\leq\pi/2$, and a corona region
$0\leq\theta<\theta_{\mathrm{d}}$ with
$\theta_{\mathrm{d}}\approx65^{\circ}$. At the beginning of the
simulation ($t=0$), there is no high density disc matter in the
domain. We fix the density and temperature in the disc region at the
outer boundary to $\rho=\rho_{\mathrm{d}}$ and $T=T_{\mathrm{d}}$.
Consequently as the simulation advances in time, a high-density
($\rho_{\mathrm{d}}$) low-temperature ($T_{\mathrm{d}}$) gas enters
the simulation domain through the disc boundary region,
$\theta>\theta_{\mathrm{d}}$. The gas in the disc region continues to
flow inward due to the $\alpha$ viscosity described earlier. The spin
of the central star is initially set to a small value corresponding to
$r_{\mathrm{cor}}=R_{\mathrm{out}}$ where
$r_{\mathrm{cor}}=(GM_{*}/\Omega^{2})^{1/3}$, and it is gradually
increased to a final stellar angular velocity, $\Omega_{*}$. As the simulation progresses,
the information about the stellar rotation propagates rapidly to the
low-density corona.

At the inner boundary ($r=R_{\mathrm{in}}$), the frozen-in condition
is applied to the poloidal component $\boldsymbol{B}_{\mathrm{p}}$ of
the field i.e., $B_{r}$ is fixed while `free' boundary conditions are
applied to $B_{\theta}$ and $B_{\phi}$ ($\partial B_{\theta}/\partial
r=0$ and $\partial B_{\phi}/\partial r=0$). The free boundary
conditions {[}$\partial\left(\cdots\right)/\partial r=0${]} are also
applied to density, pressure, and specific entropy. The velocity
components are calculated using free boundary conditions, and adjusted
to be parallel to the magnetic field vector at the inner boundary. No
matter enters from the inner boundary. The mass injection from the
inner boundary, e.g.~via a stellar wind, is not considered in this
work.

At the outer boundary ($r=R_{\mathrm{out}}$) of the corona
  region ($0\leq\theta<\theta_{\mathrm{d}}$), the free boundary
conditions are applied to all the hydrodynamic variables. The matter
is free to flow out, but no matter is allowed to enter from the outer
boundary in the coronal region. At the outer boundary of the 
  disc region, ($\theta_{d}\leq\theta\leq\pi/2$), the density is
fixed at the `disc density' ($\rho_{\mathrm{d}}$). The velocity is
fixed at a slightly sub-Keplerian velocity,
i.e. $\Omega_{\mathrm{d}}=\kappa\Omega(r_{\mathrm{d}})$ where
$\kappa=0.997$. This allows the matter to flow into the simulation
region through the boundary more easily. Finally, the boundary
conditions on the equatorial plane and on the rotation axis are
symmetric and antisymmetric, respectively.

\section{Radiative Transfer Model }

\label{sec:RT-model}

The density, velocity and modified temperature
  (Section~\ref{sub:depend-Tmax}) from a snapshot of the MHD
model, as described above, will be used as inputs for the line profile
calculations.  The hydrogen and helium line profiles from the
simulations are computed by using the radiative transfer code
\textsc{torus} (e.g.~\citealt{harries:2000}; \citealt{kurosawa:2006};
KU11). The numerical method used in this work is essentially identical
to those in KU11. Below, we briefly describe only the essential
aspects of the model, and refer readers to KU11 for more
comprehensive descriptions of our radiative transfer model.

\subsection{Outline}

The basic steps for computing the line profiles are as follows:
(1)~mapping of the density, velocity and temperature
(but see Section~\ref{sub:depend-Tmax}), from an MHD 
simulation output to the radiative transfer grid, (2)~the source
function ($S_{\nu}$) calculation and (3)~the observed flux/profile
calculation. A brief description of each step is described below.

In step (1), we use a mesh refinement grid which allows us an accurate
mapping of an MHD simulation data on to the radiative transfer grid.
In the grid construction process, we ensure that the resolution of the
radiative transfer grid is higher or equivalent to the original MHD
simulation grid. In step (2), we adopt the method described by
\citet{klein:1978} (see also \citealt{rybicki:1978};
\citealt*{hartmann:1994}) in which the Sobolev approximation method is
used. Our atomic model consists of 20 bound levels for \ion{H}{i}, 19
levels for \ion{He}{i} (up to the principal quantum number $n=4$
level), 10 levels for \ion{He}{ii}, and the continuum level for
\ion{He}{iii}. In step (3), an observed line profile will be computed
by using the source function found in step (2). The observed fluxes are 
computed in the cylindrical coordinates system whose symmetry axis
points towards an observer. The line profiles are numerically computed via
integrations of the equation of radiative transfer. 
The number of grid points used for a typical integration are
$n_{r}=180$ and $n_{\phi}=100$ for the radial 
and azimuthal grid points, respectively.  A typical number of
frequency points used across a given line profiles is
$n_{\nu}=101$. The line broadening effects (Stark and van der Waals)
are also included for some hydrogen lines (e.g.~H$\alpha$ and
H$\beta$).

\subsection{Continuum sources}

\label{sub:continum-sources}

\subsubsection{Photosphere}

\label{sub:cont-photosphere}

For all the line profile models presented in this work, we adopt stellar
parameters of a typical classical T~Tauri star for the central continuum
source, i.e.~ its stellar radius $R_{*}=2.0\,\Rsun$ and its mass
$M_{*}=0.8\,\Msun$. Thus, we adopt the model atmosphere of \citet{kurucz:1979},
with the effective temperature of photosphere $T_{\mathrm{ph}}=4,000\,\Kelvin$
and the surface gravity $\log g_{*}=3.5$ (cgs), as the photospheric
contribution to the continuum flux.

\subsubsection{Hotspots}

\label{sub:cont-hotspots}

The hot spots on CTTSs are formed by the infalling gas along the magnetic
field on to the stellar surface. As the gas approaches the surface,
it decelerates in a strong shock, and is heated to $\sim10^{6}\,\Kelvin$.
The X-ray radiation produced in the shock will be absorbed by the
gas locally, and re-emitted as optical and UV light (\citealt{calvet:1998};
\citealt{gullbring:2000}) -- forming the high temperature regions
on the stellar surface with which the magnetic field intersects. In
this study, we adopt the hot spot temperature model described by \citet{romanova:2004}
in which the local hot spot luminosity is computed directly from the
inflowing mass flux on the surface of the star from MHD simulations.
The temperature of the hot spots is determined by the conversion of
kinetic and internal energy of infalling plasma to a thermal radiation.
Using the mass flux ($\rho\, v_{r}$) at a given location on the stellar
surface, one finds the corresponding thermal radiation temperature
as  $ T_{\mathrm{hs}}=\left( \rho\,
    v_{r}\,e_{\mathrm{g}}^{2} / \sigma  \right)^{1/4} $ 
where $e_{\mathrm{g}}=\left(v^{2}/2+w\right)^{2}$. Further, $\rho$,
$v_{r}$, $v$ and $\sigma$ are the density, the radial component of
velocity, the speed of plasma, and the Stefan-Boltzmann constant,
respectively. The specific enthalpy of the gas is
$w=\gamma\left(p/\rho\right)\left(\gamma-1\right)$ where $\gamma$ is
the adiabatic index and $p$ is the gas pressure.  We compare this
temperature $T_{\mathrm{hs}}$ with the effective temperature of
photosphere $T_{\mathrm{ph}}$ to determine the shape and the size of
hot spots. When $T_{\mathrm{hs}}>T_{\mathrm{ph}}$, the location on the
stellar surface is flagged as hot. For the hot surface, the total
continuum flux is the sum of the blackbody radiation with
$T_{\mathrm{hs}}$ and the flux from the model photosphere mentioned
above. The contribution from the inflow gas is ignored when
$T_{\mathrm{hs}}<T_{\mathrm{ph}}$.

\subsubsection{X-ray emission}

\label{sub:cont-x-ray}

Relatively strong X-ray emission is commonly found in CTTSs, with
their luminosities $L_{\mathrm{X}}\left(0.3-10\,\mathrm{keV}\right)$
ranging from $\sim10^{28}\,\mathrm{erg\, s^{-1}}$ to $\sim10^{31}\,\mathrm{erg\, s^{-1}}$
(e.g.~\citealt{Telleschi:2007}; \citealt{Gudel:2007}; \citealt{Gudel:2010}).
Recent studies by \citet{Kwan:2011} and KU11 have shown that the
photoionization by high energy photons, in addition to the normal
continuum emission from the photosphere, is most likely needed 
for generating the high opacities and hence strong absorption features
seen in the observed \ion{He}{i}~$\lambda$10830 line profiles of
CTTSs (e.g.~ED06). As in KU11, we adopt a simplest
model of the X-ray radiation here. We assume that the X-ray radiation
arises uniformly from the stellar surface as though it were formed
in the chromosphere. This allows us to include the X-ray emission
by simply adding it to the normal stellar continuum flux (Section~\ref{sub:cont-photosphere}).
The dependency of the line formation on the different X-ray emission
locations is beyond the scope of this paper, but will be explored
in the future. In this work, we assume that the X-ray emitting plasma
is radiating thermally (as a blackbody) with a single temperature
$T_{\mathrm{X}}$. The thermal X-ray radiation flux is normalised
with the total X-ray luminosity $L_{\mathrm{X}}\left(0.1-10\,\mathrm{keV}\right)$,
which we set as an input parameter. 
Note that we restrict the contribution of the X-ray emitting gas to
the continuum to be only within the X-ray energy range (0.1--10~keV),
i.e.\,the intensity ($I_{\nu}$) from the X-ray emitting plasma is assumed
to be zero for the energy below 0.1~keV and above 10~keV.
The attenuation of X-ray flux
is not taken into account in our model; however, this is to be improved
in a future study.
Further discussion on the X-ray optical depth can be found at the end
of Section~\ref{sub:depend-X-ray}.  

Note that the stellar photospheric continuum in our model 
(Section~\ref{sub:cont-photosphere}) has a relatively low effective
temperature ($4,000\,\Kelvin$); therefore, the stellar
flux in EUV (including that at 24.58~eV) is essentially zero. 
This is situation could be also thought as the case when the EUV flux
is completely attenuated. With no EUV flux present, the only source of
the photoionization of \ion{He}{i} in our model is the X-ray emission as
described above.


\begin{table*}

\caption{Summary of MHD Models}

\label{tab:MHD-Model-Summary}

\begin{tabular}{rcccccccc}
\hline 
 & 
$T_{\mathrm{d}}$
 & 
$T_{\mathrm{c}}$
 & $\rho_{\mathrm{d}}$ & $\rho_{\mathrm{c}}$ & 
$\dot{M}_{\mathrm{a}}$
 & 
$\dot{M}_{\mathrm{cw}}$
 & 
$\dot{M}_{\mathrm{pw}}$
 & 
Funnel Flow Size
\tabularnewline

Model ID
 & 
{(}K{)}
 & 
{(}K{)}
 & 
{(}$\mathrm{g\, cm^{-3}}${)}
 & 
{(}$\mathrm{g\, cm^{-3}}${)}
 & 
{(}$\MsunPerYear${)}
 & 
{(}$\MsunPerYear${)}
 & 
{(}$\MsunPerYear${)}
 & 
{(}$R_{*}${)}
\tabularnewline
\hline 

A
 & 
$1.5\times10^{3}$
 & 
$9.2\times10^{6}$
 & 
$4.9\times10^{-12}$
 & 
$8.2\times10^{-16}$
 & 
$3.4\times10^{-7}$
 & 
$8.2\times10^{-8}$
 & 
$1.4\times10^{-10}$
 & 
$\sim1.2$
\tabularnewline

B
 & 
$1.5\times10^{3}$
 & 
$9.2\times10^{6}$
 & 
$8.2\times10^{-13}$
 & 
$1.4\times10^{-16}$
 & 
$4.1\times10^{-8}$
 & 
$6.9\times10^{-9}$
 & 
$2.6\times10^{-10}$
 & 
$\sim2.0$
\tabularnewline
\hline
\end{tabular}

\end{table*}



\begin{figure*}

\begin{center}
\begin{tabular}{ccc}
  \includegraphics[clip,width=0.31\textwidth]{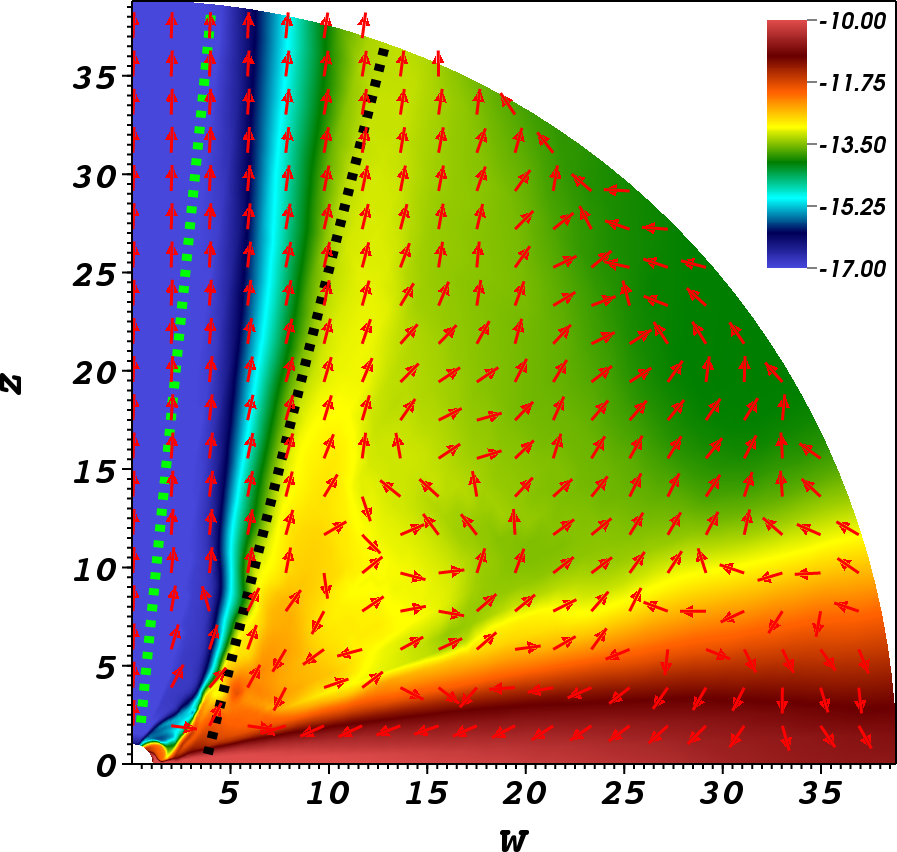} & 
  \includegraphics[clip,width=0.31\textwidth]{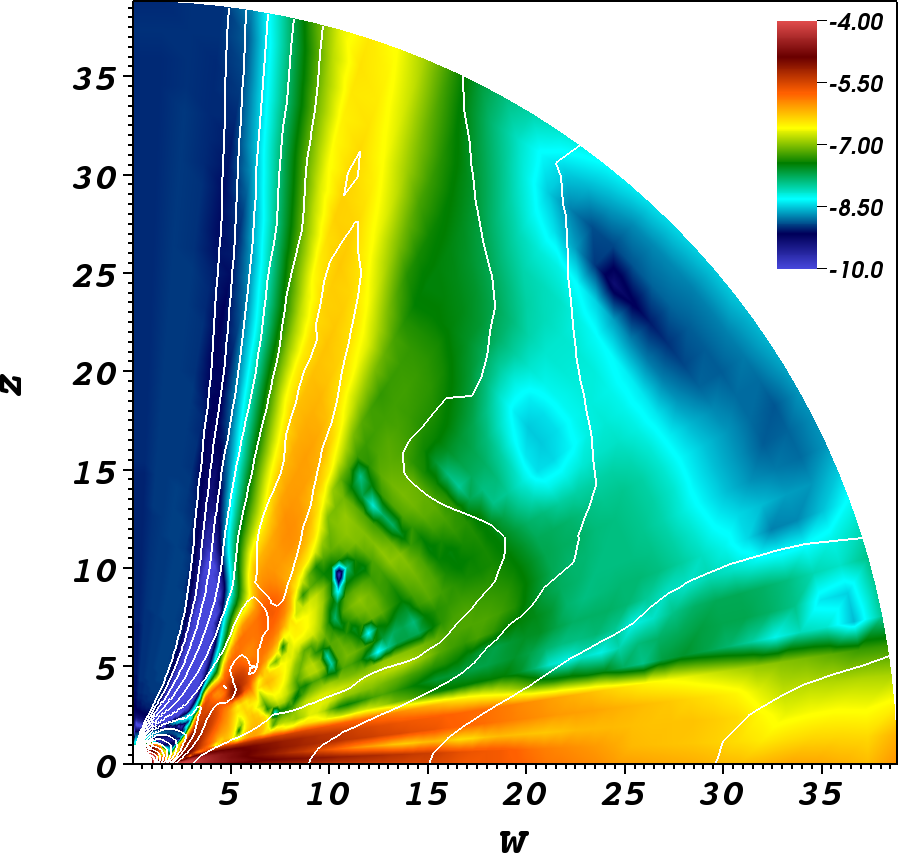} & 
  \includegraphics[clip,width=0.335\textwidth]{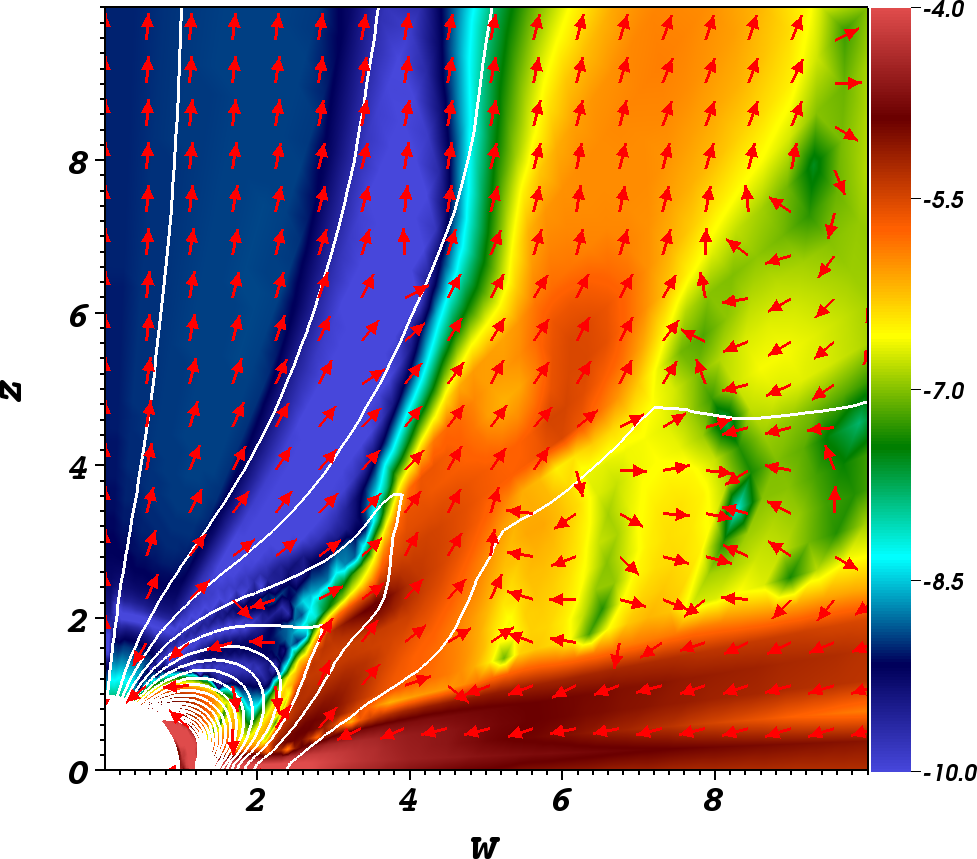}
  \tabularnewline
  \includegraphics[clip,width=0.31\textwidth]{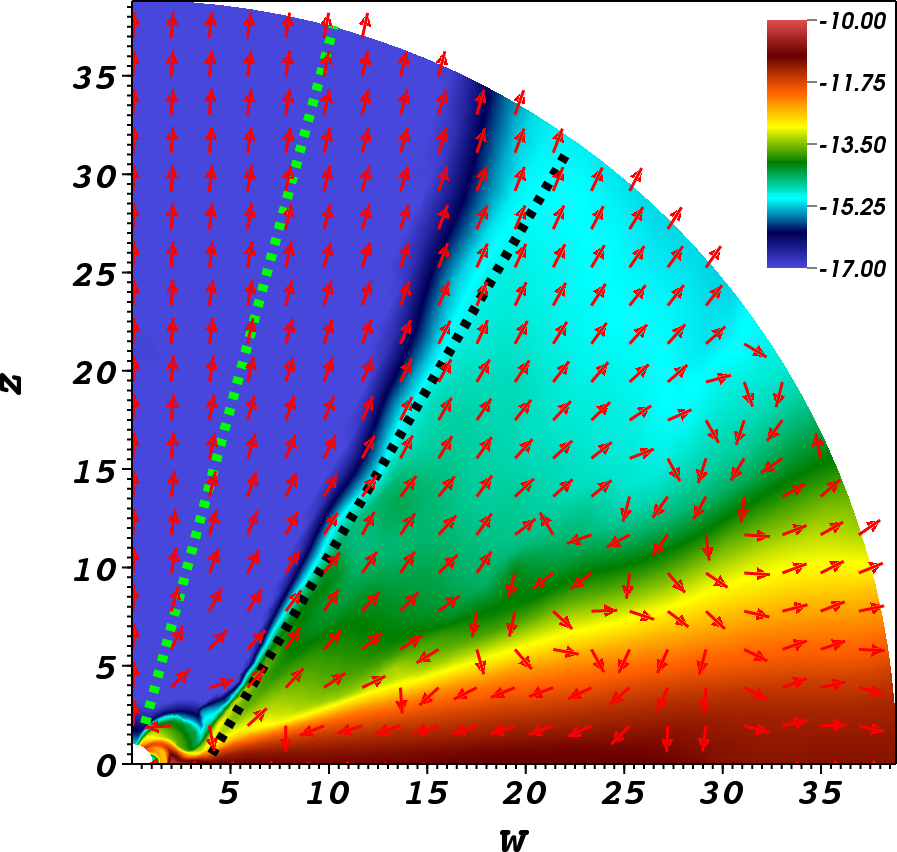} & 
  \includegraphics[clip,width=0.31\textwidth]{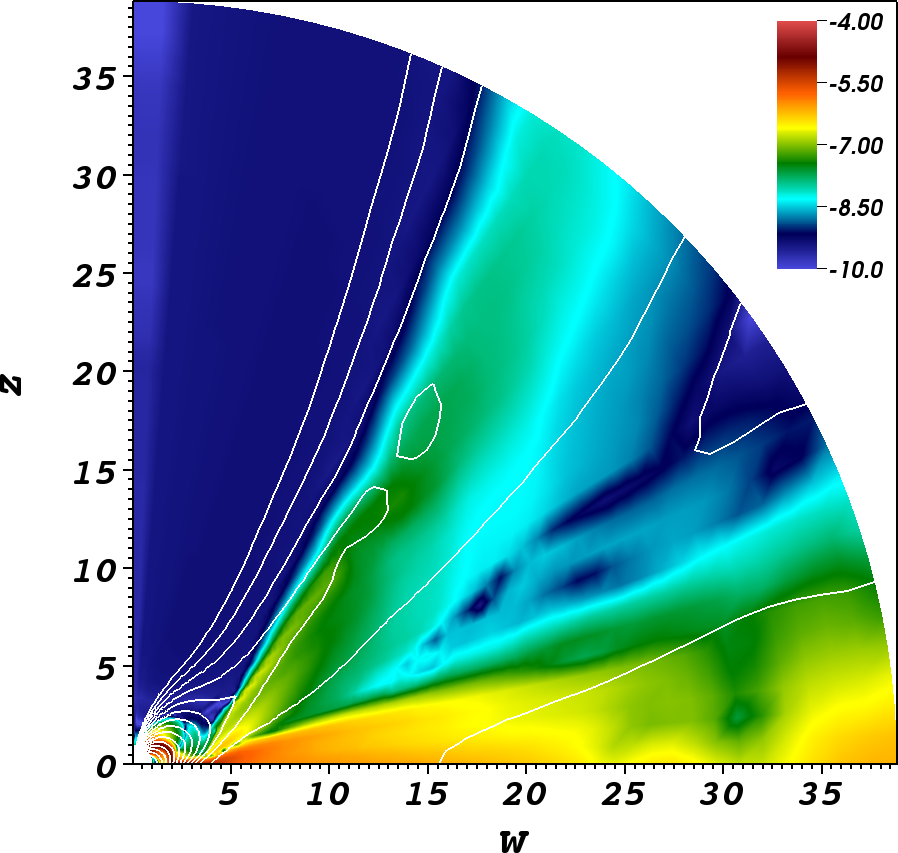} & 
  \includegraphics[clip,width=0.335\textwidth]{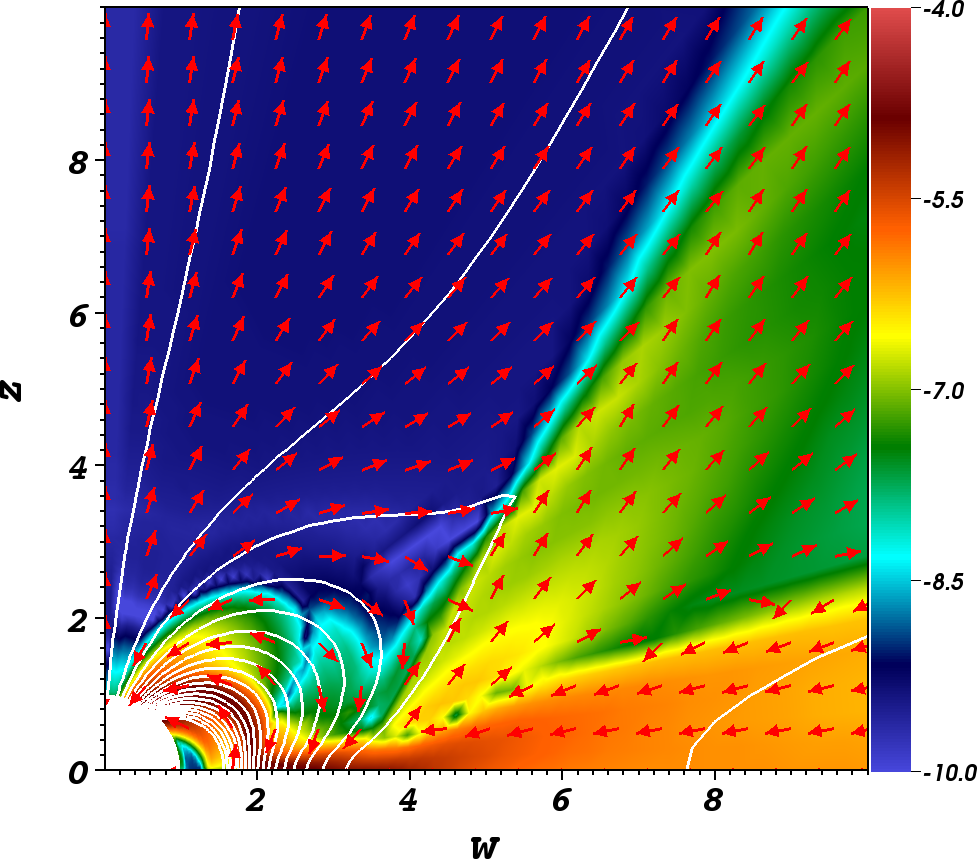}
  \tabularnewline
\end{tabular} 
\end{center}

\caption{Summary of MHD models with a high mass-accretion rate
  (Models~A; \emph{top panels}) and a moderate mass-accretion rate (Model
  B; \emph{bottom panels}). See Table~\ref{tab:MHD-Model-Summary} for the
  model parameters. The density maps (in logarithmic scale; in cgs
  units) are over-plotted with the directions of the poloidal velocity
  ($v_{p}$) as arrows (\emph{left panels}). The poloidal matter flux
  $v_{p}\,\rho$ maps (in logarithmic scale; in cgs units) are shown
  with sample magnetic field lines (\emph{middle panels}). The zoom-in
  poloidal matter flux $v_{p}\,\rho$ maps (\emph{right panels}) are
  shown with both sample magnetic field lines and poloidal velocity
  directions as arrows. In the left panels, the reference lines
  passing through the conical-shell (\emph{black dashed line}) and  
  polar winds (\emph{green dashed line}) are also shown. The length
  scales are in the units of the stellar radius ($R_{*}$). } 

\label{fig:MHD01}

\end{figure*}



\begin{figure*}

\begin{center}
\begin{tabular}{cc}
  \includegraphics[clip,width=0.4\textwidth]{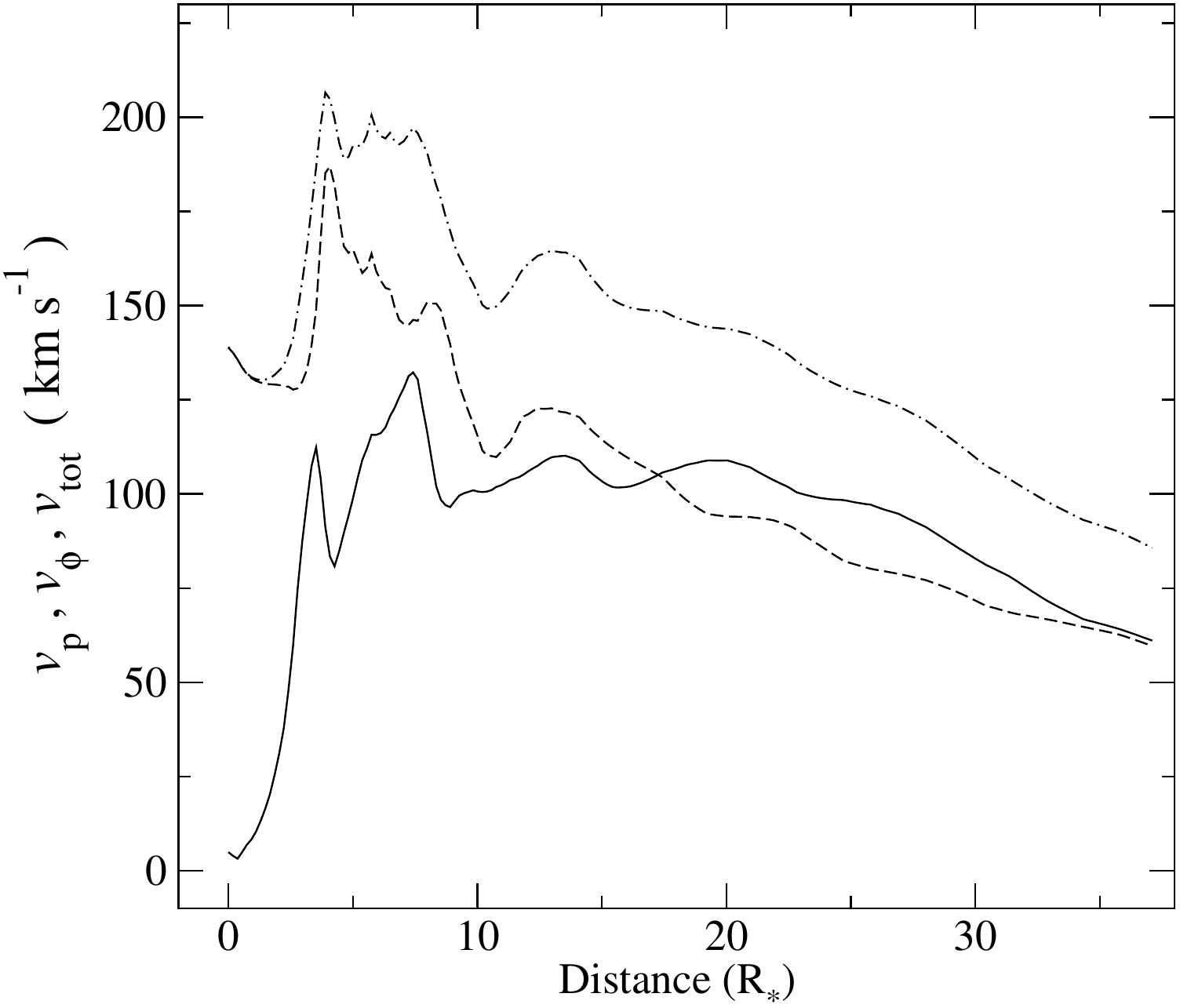} & 
  \includegraphics[clip,width=0.4\textwidth]{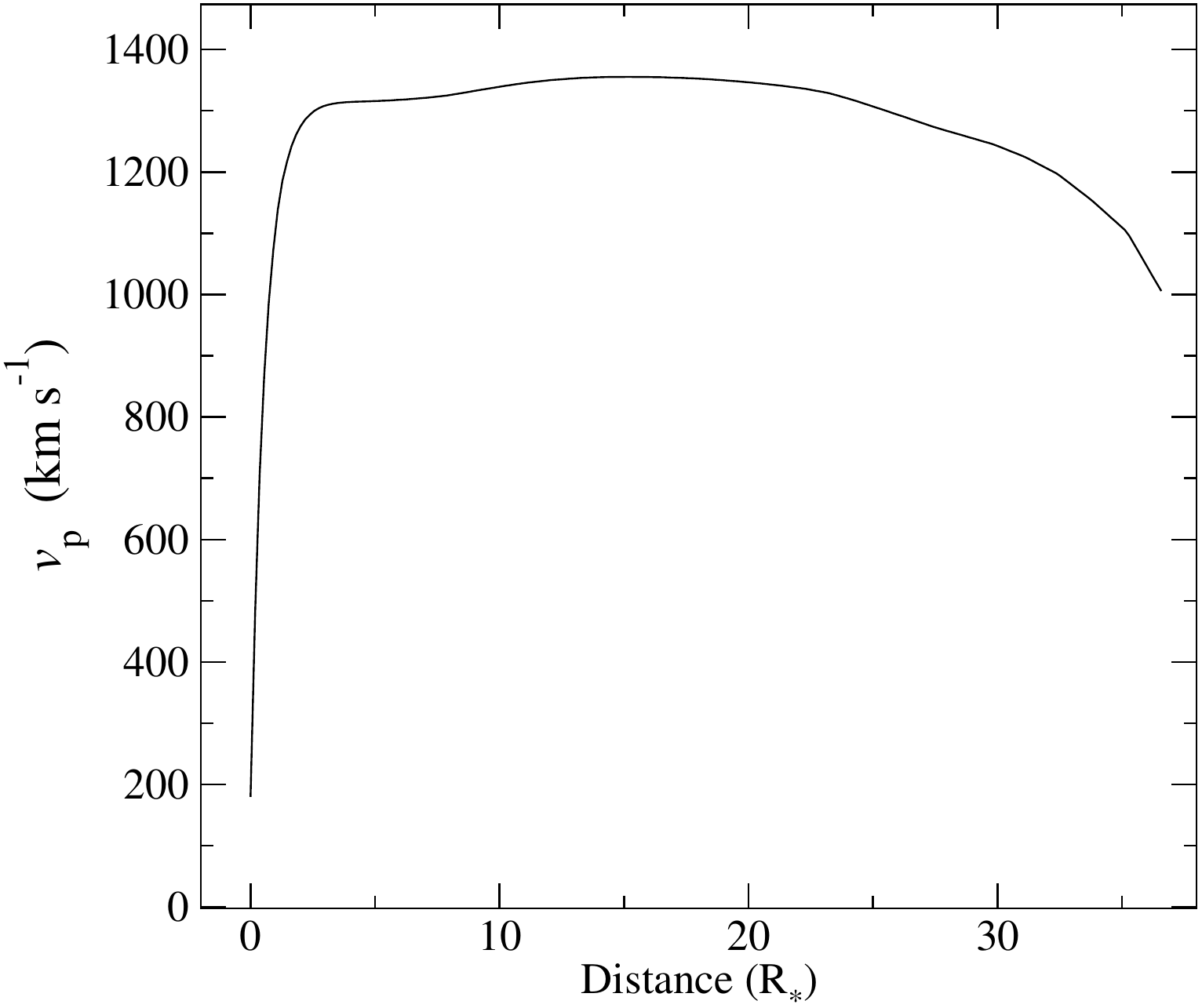}
  \tabularnewline
  \includegraphics[clip,width=0.4\textwidth]{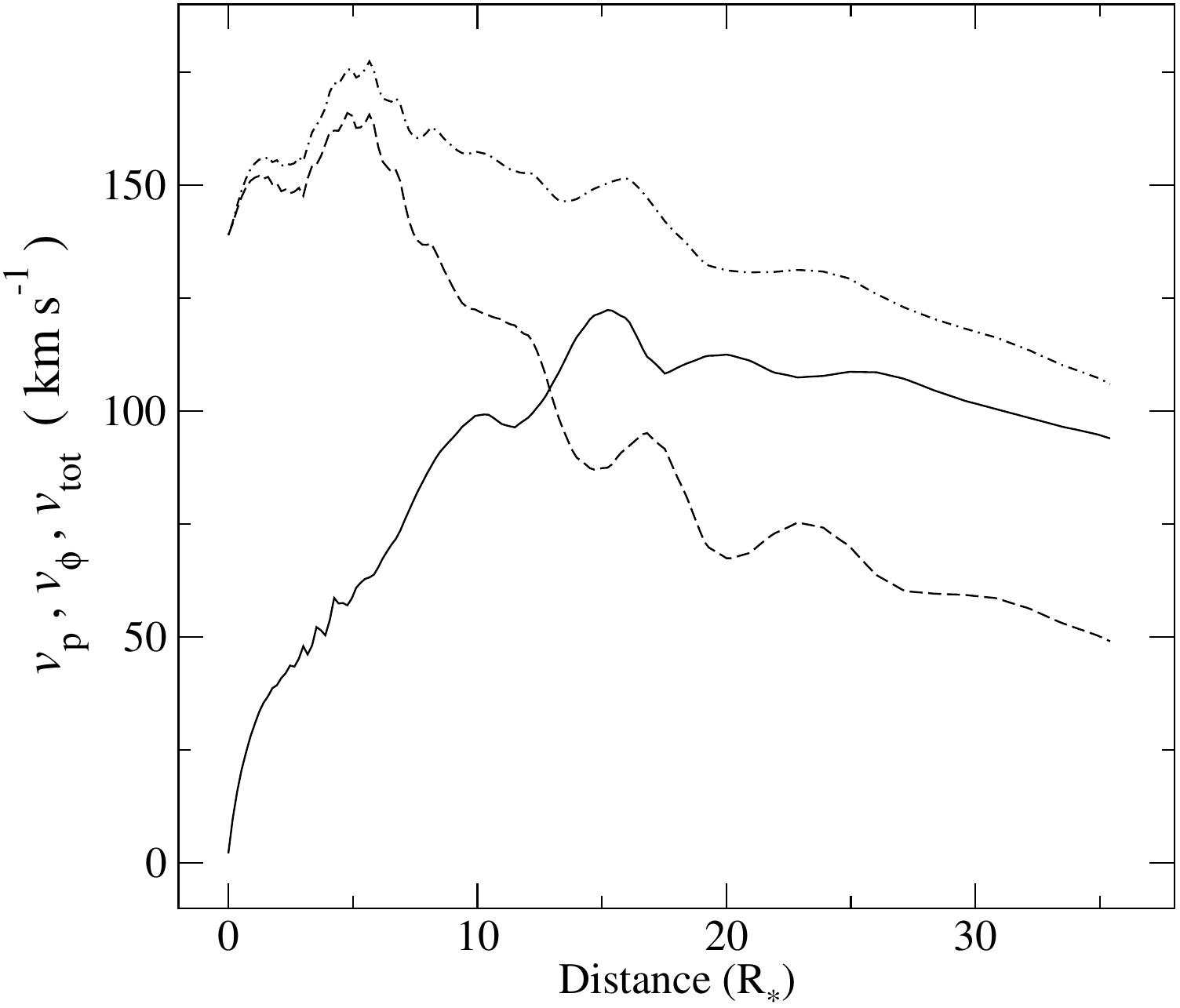}  & 
  \includegraphics[clip,width=0.4\textwidth]{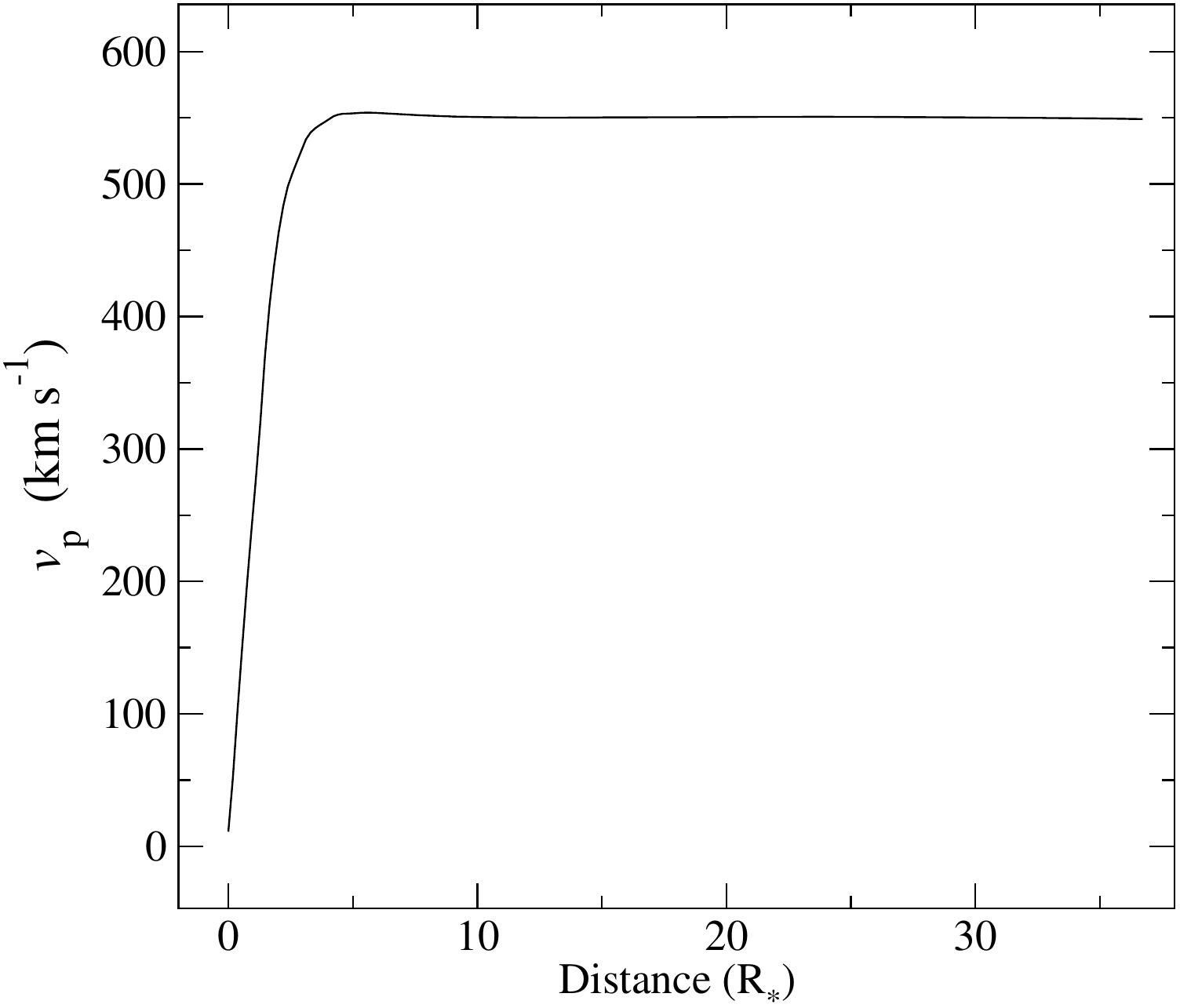}
  \tabularnewline
\end{tabular}
\par\end{center}

\caption{The total speed ($v_{\mathrm{tot}}$; \emph{dash-dot}), the
  poloidal ($v_{p}$; \emph{solid}) and azimuthal components
  ($v_{\phi}$; \emph{dash}) of the velocity along the reference lines
  indicated in the left panels of Fig.~\ref{fig:MHD01} are shown for Models~A (\emph{top
    panels}) and B (\emph{bottom panels}).  The horizontal axis
  indicates the distance ($d$) from the base of the wind along the
  reference lines.  The panels on the \emph{left column} are the
  velocity components along the reference line that goes through the
  `conical-shell winds.'  The panels on the \emph{right column} show
  $v_{\mathrm{p}}$ along the reference line that go through 
  'the polar wind'. Compared to $v_{\mathrm{p}}$, $v_{\phi}$ is
  negligibly small and $v_{\mathrm{p}}\approx v_{\mathrm{tot}}$ in the
  polar wind; hence, they are omitted from the plots.  }

\label{fig:MHD-Vel}

\end{figure*}

\section{MHD model results}

\label{sec:mhd-results}

\subsection{Input parameters and grid setup}

\label{sub:mhd-setup}

The stellar parameters adopted here are $M_{*}=0.8\, M_{\odot}$,
$R_{*}=2.0\, R_{\odot}$ and $P=5.38$~d where $M_{*}$, $R_{*}$
and $P$ are the stellar mass, radius and the rotational period, respectively.
The corresponding corotation radius and the
escape velocity of the system are $6.0\, R_{*}$ and $390\,\kmps$,
respectively. Following, RO09, we set the dimensionless viscosity
and diffusivity coefficients (see Section~\ref{sub:mhd-equations})
to $\alpha_{\mathrm{v}}=0.3$ and $\alpha_{\mathrm{d}}=0.1$. For
the models presented here, the surface magnetic field strength is
fixed at $B_{*}=1$\,kG, and the magnitude of the dipole moment (equation~\ref{eq:B-dipole})
at $\mu=40$ in the units of $B_{*}$. 
The initial corona temperature is set as
$T_{\mathrm{c}}=9.2\times10^{6}$~K, and the disc temperature at the
outer boundary is fixed at  $T_{\mathrm{d}}=1.5\times10^{3}$~K  (see
Section~\ref{sub:mhd-equations}). The ratio of the corona to disc
densities ($\rho_{\mathrm{c}}/\rho_{\mathrm{c}}$) are fixed at
$\sim1.7\times10^{-4}$, and we vary $\rho_{\mathrm{d}}$ to control the
disc mass-accretion rate. 

The simulations are performed in the radial range, $R_{\mathrm{in}}\leq r\leq R_{\mathrm{out}}$
where $R_{\mathrm{in}}=1.0\, R_{*}$ and $R_{\mathrm{out}}=33.7\, R_{\mathrm{in}}$,
and in the polar angle range, $0\leq\theta\leq\pi/2$. The grid spacing
in $\theta$ is uniform, but that in $r$ is chosen such that each
side of the curvilinear rectangle cells is approximately equal. The
number of grid points used in the models are $n_{\theta}=40$ and
$n_{r}=96$. Note that the grid setup used here is slightly different
from those in the conical-shell wind models of RO09 who used the radial
range: $R_{\mathrm{in}}=2.0\, R_{*}$ and $R_{\mathrm{out}}=16\, R_{\mathrm{in}}$,
and the number of grid points: $n_{\theta}=31$ and $n_{r}=51$. Their
corotation radius for a CTTS model is $3.0\, R_{*}$
which is 1/2 of our value. We have run simulations with higher resolutions
(e.g.~$n_{\theta}=80$ and $n_{r}=192$), but overall flow patterns
as well as the physical quantities (e.g.~plasma temperature, density
and velocity) are very similar to those of the lower resolution models
presented in this work.

\subsection{Matter flow in conical-shell winds}

\label{sub:mhd-main-results}

We have performed the MHD simulations with ranges of the disc density
$\rho_{\mathrm{d}}$ and have examined the dependency of the flows
on $\rho_{\mathrm{d}}$. Here, we present two most representative
cases for (1)~a relatively high mass-accretion rate (Model~A) and
(2)~a moderate mass-accretion rate (Model~B). Fig.~\ref{fig:MHD01}
shows the maps of density ($\rho$), poloidal mass-flux ($\Phi_{\mathrm{m}}\equiv\rho v_{\mathrm{p}}$)
along with sample magnetic field lines and the directions of the poloidal
velocity ($v_{\mathrm{p}}$), for both models. The main model parameters
and the corresponding mass-accretion rates ($\dot{M}_{\mathrm{a}}$),
mass-loss rates in the conical-shell wind ($\dot{M}_{\mathrm{cw}}$) and
the mass-loss rates in the polar wind ($\dot{M}_{\mathrm{pw}}$)
are summarised in Tab.~\ref{tab:MHD-Model-Summary}. In the following,
each model will be described in more detail.

\subsubsection{A relatively high mass-accretion rate case: Model~A }

\label{sub:MHD-Model-A}

This model uses a relatively large disc density (at the outer boundary),
$\rho_{\mathrm{d}}=4.9\times10^{-12}\,\mathrm{g\, cm^{-3}}$. The
disc matter initially entering from the outer boundary forms an accretion
disc around the star, and reaches a semi-steady state as shown in
Fig.~\ref{fig:MHD01}. The high disc density matter pushes the stellar
dipole magnetic fields almost all the way to the stellar surface,
and a very small funnel ($\sim1.2\, R_{*}$ in size) is formed.
The matter accretes on to the stellar surface through the small funnel
with the accretion rate $\dot{M}_{\mathrm{a}}=3.4\times10^{-7}\,\MsunPerYear$.
The geometrically inclined field lines near $r\approx2\, R_{*}$ on the
disc plane and the inflated 
field above it create conditions favourable for matter from the inner
disc region (RO09). The gas flows, from the inner disc region, into
a conical-shell shaped wind. The flow is somewhat collimated near
the outer boundary and has a half-opening angle $\theta_{\mathrm{o}}\approx18^{\circ}$.
This is smaller than those of the conical-shell wind simulations of RO09
($\theta_{\mathrm{o}}\sim30^{\circ}-40^{\circ}$) who used a smaller
radial range ($R_{\mathrm{out}}=16\, R_{\mathrm{in}}$) than our value
($R_{\mathrm{out}}=33.7\, R_{\mathrm{in}}$). On the other hand, \citet{Lii:2012}
obtained a much more collimated outflow ($\theta_{\mathrm{o}}\approx4^{\circ}$)
using a larger radial range ($R_{\mathrm{out}}=42\, R_{\mathrm{in}}$). 

The poloidal and toroidal velocity components ($v_{\mathrm{p}}$ and
$v_{\phi}$) along with the magnitude of the total velocity ($v_{\mathrm{tot}}$)
along the conical-shell wind (along the black dashed line in the upper right panel
of Fig.~\ref{fig:MHD01}) are shown in Fig.~\ref{fig:MHD-Vel}.
The rotational speed $v_{\phi}$ dominates near the base of the wind,
and its value is essentially the same as that of the Keplerian velocity
at the wind base. The poloidal component $v_{\mathrm{p}}$ increases
along the conical-shell wind, and it peaks around $d\approx8\, R_{*}$ with
$v_{\mathrm{p}}\approx125\,\kmps$, then it slowly decreases
beyond that point, and $v_{\mathrm{p}}$ reaches $\sim60\,\kmps$ 
when the conical-shell wind escapes from the outer boundary. 
The slight decrease in $v_{\mathrm{p}}$ at large radii is
related to the fact that the reference lines in Fig.~\ref{fig:MHD01}
are not exactly parallel to the outflow stream lines, and
$v_{\mathrm{p}}$ tends to decrease as the distance from the symmetry
axis ($z$) increases at a given value of $z$.  
The velocity range found here is consistent with the extent of the
narrow blueshifted absorption components (up to $\sim200\,\kmps$) in
the observed \ion{He}{i}~$\lambda10830$ profiles obtained by
e.g.~ED06. 

Fig.~\ref{fig:MHD01} also shows the lower density and higher speed
outflow in the polar direction which we refer to as `the polar wind.'
This should be distinguished from `the stellar wind' which arises from/near
the stellar surface, but is not implemented in this model. The matter
in the polar wind is simply the gas redirected from the upper part
of the accretion funnel flow, and it does not originate from the stellar
surface. The poloidal component of the velocity in the polar wind (along
the green dashed line in the upper right panel of Fig.~\ref{fig:MHD01})
is also shown in Fig.~\ref{fig:MHD-Vel}. The polar wind quickly
accelerates to about $1.3 \times 10^3$~$\kmps$ in a few stellar radii, and
it slowly decelerates beyond $d\approx15\, R_{*}$.  
The density in the polar wind is very low and the polar region is
magnetically dominated. As a result, the low density gas is  
accelerated up to a very high velocity via strong magnetic forces. 
A similar component of the outflow has been observed in the
simulations for the propeller regime (a fast stellar rotation
regime) (e.g.~\citealt{Ustyugova:2006}; \citealt{Romanova:2009}). 
As mentioned before, the slight decrease in $v_{\mathrm{p}}$ at large
radii seen in Fig.~\ref{fig:MHD-Vel} is
related to the fact that the reference lines in Fig.~\ref{fig:MHD01}
are not exactly parallel to the outflow stream lines.
This velocity is much larger than the extent of the wide and deep blueshifted absorption
feature ($\sim400\,\kmps$) in the observed \ion{He}{i}~$\lambda10830$
profiles (e.g.~ED06). The poloidal velocity of
the polar wind is about 10 times larger than that of the conical-shell wind
in most of the radial ranges. The mass-loss rates in the conical-shell and
polar winds are $\dot{M}_{\mathrm{cw}}=8.2\times10^{-8}\MsunPerYear$
and $\dot{M}_{\mathrm{pw}}=1.4\times10^{-10}\MsunPerYear$, respectively.
The former is much larger and the latter is smaller than a typical
mass-loss rate of CTTSs ($\sim10^{-9}\,\MsunPerYear$), but they are
still within the range of the observed mass-loss rates (e.g.~\citealt{hartigan:1995}). 

Although the mass-accretion rate
$\dot{M}_{\mathrm{a}}=3.4\times10^{-7}\,\MsunPerYear$ found in Model~A
is much higher than that of a typical CTTS,
$\sim10^{-8}\,\MsunPerYear$(e.g.~\citealt{gullbring:1998};
\citealt{Hartmann:1998}; \citealt*{Calvet:2000}), it is still an
acceptable value for CTTSs (e.g.~\citealt{hartigan:1995}). However,
this model is not suitable for modelling line profile for CTTSs
because the size of the magnetospheric accretion funnel flow is much
smaller than that indicated by previous studies
(e.g.~\citealt*{Muzerolle:1998b}; \citealt*{muzerolle:2001}). The
observed line profiles (e.g.~ Pa$\beta$, Br$\gamma$ and
\ion{He}{i}~$\lambda10830$) often show redshifted absorption features
that often extends to 200--300~$\kmps$ (e.g.~\citealt{edwards:1994};
\citealt{alencar:2000}; \citealt{folha:2001}) which are formed in the
inflowing gas that is accreting on to the stellar surface through the
funnel flows. When the size of the magnetosphere or the funnel flow is
too small, the inflow speed in the funnel also becomes too small since
the gas gains the kinetic energy from the gravitational potential
energy (c.f.~\citealt*{ghosh:1977}; \citealt{hartmann:1994}).  In
Model~A, we find that the poloidal velocity of the gas in the funnel
flow is only $\sim120\,\kmps$ which is much smaller than a typical
observed value (200--300~$\kmps$). Further, the geometry of the funnel
flow is not ideal for producing the redshifted absorption
feature. Since the funnel flow is located very close to the stellar
surface, the funnel stream is almost parallel to the stellar surface.
This flow geometry reduces the line-of-sight (recession) speed of the
gas toward an observer. The line profile models based on this type of
flow will result in a very weak redshifted absorption component with a
velocity extent that is too small when compared to observations. To improve
the MHD model to be more suitable for CTTSs, we now attempt to enlarge
the size of the accretion funnel by preventing the accretion disc from
pushing the stellar dipole magnetic field too close to star.

\subsubsection{A moderate mass-accretion rate case: Model~B}

\label{sub:MHD-Model-B}

To reduce the amount of the compression of the magnetosphere towards
the stellar surface by the accretion disc, the disc density
$\rho_{\mathrm{\mathrm{d}}}$ at outer boundary is now reduced by a
factor of $\sim6$ from the value used in Model~A. The lower density
plasma reduces the total gravitational force on the magnetic field
lines, and consequently the size of the magnetosphere or the accretion
funnel becomes much larger ($\sim2.0\, R_{*}$) than the pervious case,
Model~A. In Model~B, the funnel flow is clearly detached from the stellar surface
(Fig.~\ref{fig:MHD01}).  The corresponding mass accretion rate on to
the stellar surface through the funnel is
$\dot{M}_{\mathrm{a}}=4.1\times10^{-8}\,\MsunPerYear$, which is very
similar to the value found for a typical CTTS
($\sim10^{-8}\,\MsunPerYear$, e.g.~\citealt{gullbring:1998};
\citealt{Hartmann:1998}; \citealt{Calvet:2000}), and is about $\sim8$
times smaller than that of Model~A. The inner edge of the accretion
disc is now located $\sim2.0\, R_{*}$ from the origin, and the outflow
in the conical-shell wind originates at a larger distance $r\approx4.0\,
R_{*}$ (in the disc plane) which is slightly smaller than the
corotation radius ($R_{\mathrm{cr}}=6.0\, R_{*}$).

The outflow geometry is similar to that of Model~A, but the wind is
less collimated with its half-opening angle of the conical-shell wind
(measured at the outer boundary)
$\theta_{\mathrm{o}}\approx35^{\circ}$ (Fig.~\ref{fig:MHD01}), which
is similar to the value found in RO09.  Fig.~\ref{fig:MHD-Vel} also
shows the poloidal and toroidal velocity components ($v_{\mathrm{p}}$
and $v_{\phi}$), and the magnitude of the total velocity
($v_{\mathrm{tot}}$) along the conical-shell wind (along the black dashed line in
the lower right panel of Fig.~\ref{fig:MHD01}).  As also seen in
Model~A, the rotational speed $v_{\phi}$ dominates near the base of the
wind. The poloidal component $v_{\mathrm{p}}$ increases along the
conical-shell wind, and it peaks around $d\approx15\, R_{\odot}$ with
$v_{\mathrm{p}}\approx125\,\kmps$. The flow then slowly decelerates
beyond that point, and $v_{\mathrm{p}}$ reaches $\sim90\,\kmps$ near
the outer boundary. As in the previous model (Model~A), the slight
decrease in $v_{\mathrm{p}}$ at large radii is 
related to the fact that the reference lines in Fig.~\ref{fig:MHD01}
are not exactly parallel to the outflow stream lines. 

Again, this velocity range in the conical-shell wind is
consistent with the velocity range of the narrow blueshifted
absorption components (up to $\sim200\,\kmps$) found in the observed
\ion{He}{i}~$\lambda10830$ profiles
(e.g.~ED06). Note that the outflow speed
($v_{\mathrm{p}}$) in the conical-shell wind in Model~B is similar to that
of Model A. The polar wind quickly accelerates to $\sim550\,\kmps$
(along the green dashed line in the lower right panel of 
Fig.~\ref{fig:MHD01}) at the distance $d\approx5\, R_{*}$ from the
base of the wind. The wind speed then slightly decreases, but it
remains almost constant all the way to the outer boundary. The outflow
speed ($v_{\mathrm{p}}$) of the polar wind in Model~B is about 2 times
less than that of Model~A.  This velocity is comparable to the extent
of the wide and deep blueshifted absorption feature ($\sim400\,\kmps$)
in the observed \ion{He}{i}~$\lambda10830$ profiles
(e.g.~ED06). The mass-loss rates in the conical-shell and
polar winds are $\dot{M}_{\mathrm{cw}}=6.9\times10^{-9}\MsunPerYear$
and $\dot{M}_{\mathrm{pw}}=2.6\times10^{-10}\MsunPerYear$,
respectively.  The former is larger and the latter is smaller than a
typical mass-loss rate of CTTSs ($\sim10^{-9}\,\MsunPerYear$), but
they are still within the range of the observed mass-loss rates
(e.g.~\citealt{hartigan:1995}).

The geometry of the well defined accretion funnel in Model~B is very
similar to that of the magnetospheric accretion funnels used in
hydrogen line profiles models e.g.~\citet{hartmann:1994},
\citet{muzerolle:2001}, \citet*{symington:2005} and
\citet{kurosawa:2006} who adopted the axisymmetric magnetosphere model
of \citet{ghosh:1977} (see also \citealt{ghosh:1979};
\citealt{Ghosh:1979b}). These models have been successfully
demonstrated that the redshifted absorption features or the inverse
P-Cygni profiles seen in some of the observed hydrogen lines can be
explained by the absorption by the inflowing gas through the
magnetospheric accretion funnels. In Model~B, we find that the flow
speed in the funnel near the stellar surface reaches $\sim250\,\kmps$,
which is very similar to the values seen in the observation,
200--300~$\kmps$ (e.g.~\citealt{edwards:1994};
\citealt*{Muzerolle:1998a}; \citealt{alencar:2000}; 
\citealt{folha:2001}).

In summary, Model~B shows very similar inflow and outflow velocities,
mass-loss and mass-accretion rates to those found in the
observations. Hence, it is a very good candidate for being tested with
our radiative transfer model which uses the MHD simulation results as
model inputs. In the following, we describe the hydrogen and helium
line profile models based on the flow solution found in Model~B.

\section{Line Profile Models}

\label{sec:results-profiles}

\begin{figure*}

\begin{center}
\begin{tabular}{cc}

  \includegraphics[clip,width=0.355\textwidth]{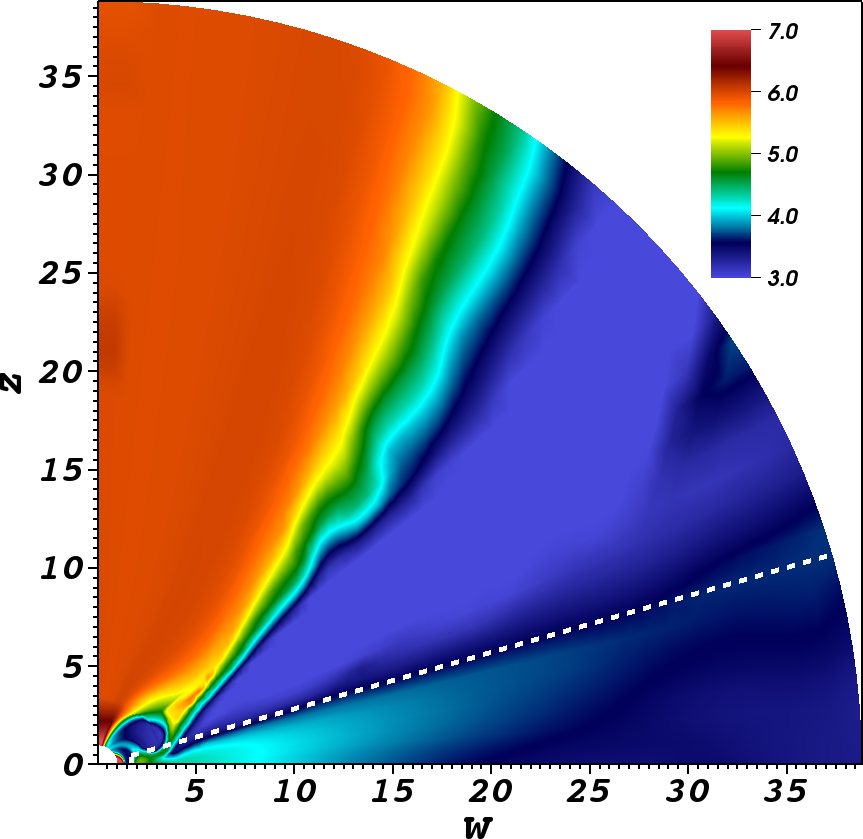} &
  \hspace{1cm}
  \includegraphics[clip,width=0.38\textwidth]{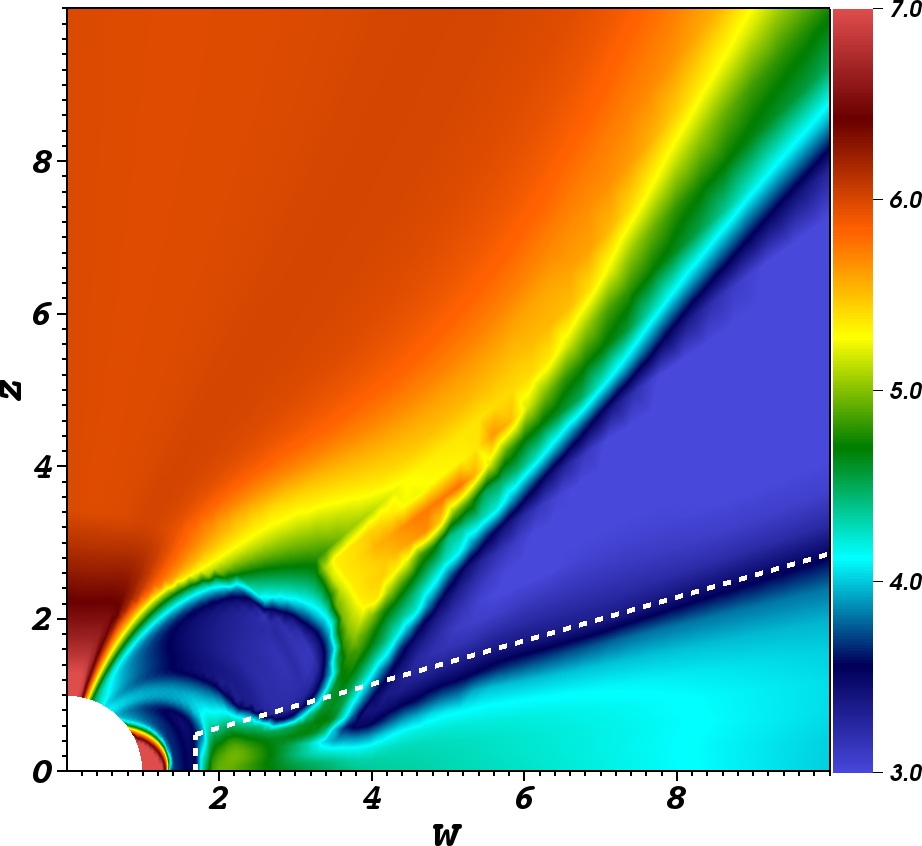} 
  \tabularnewline
  \includegraphics[clip,width=0.355\textwidth]{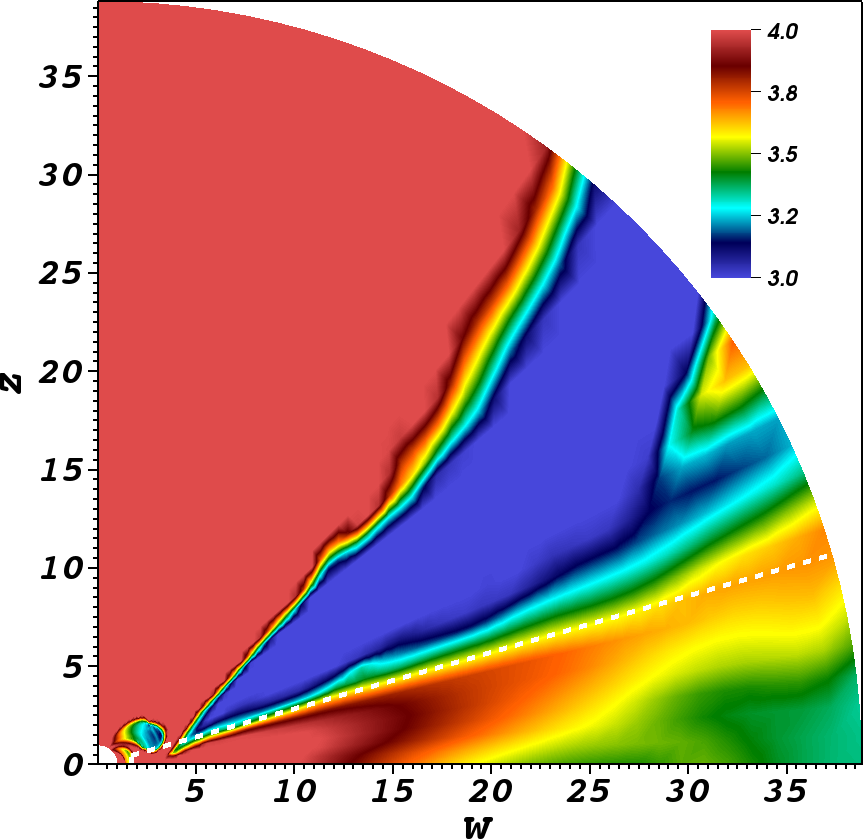} &
  \hspace{1cm}
  \includegraphics[clip,width=0.38\textwidth]{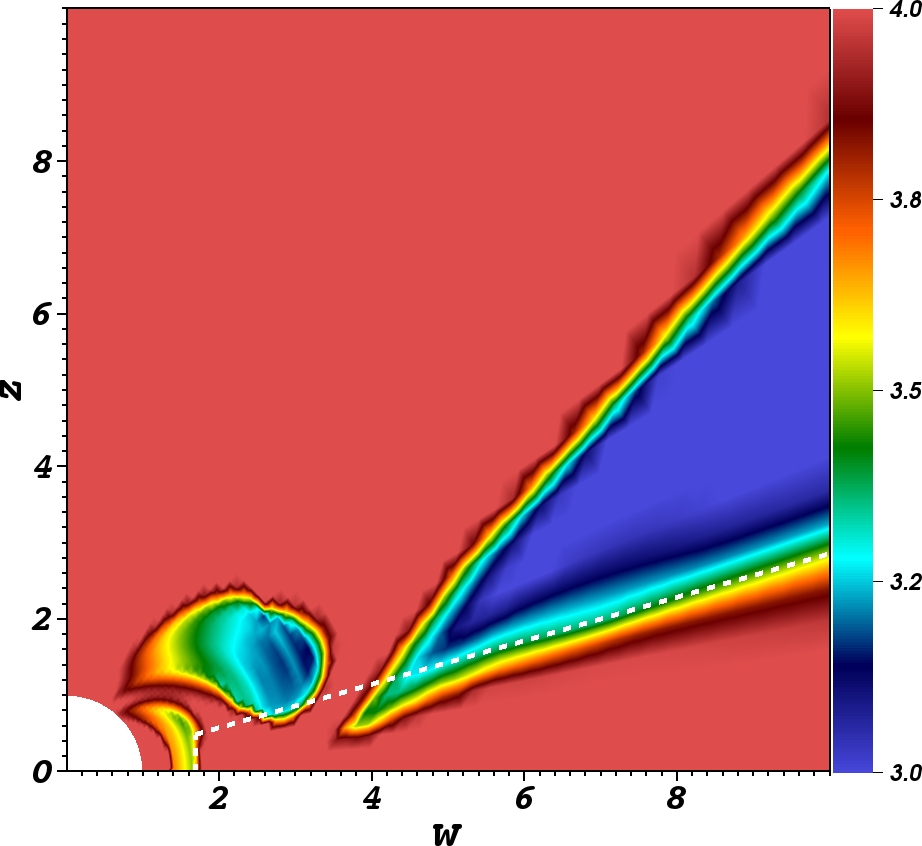} 
  \tabularnewline

\end{tabular}
\par\end{center}

\caption{ 
  The temperature maps (in logarithmic scale; in $\Kelvin$) of
  MHD Model~B (cf. Table~\ref{tab:MHD-Model-Summary} and 
  Fig.~\ref{fig:MHD01}) in full-scale (\emph{left panels})
  and in small-scale (\emph{right panels}). The \emph{upper panels}
  show the original temperatures (unmodified) from the MHD
  simulation. The \emph{lower panels} show the temperatures from
  the same MHD simulation, but the upper limit of the temperature is
  set to  $T_{\mathrm{max}}=10^{4}\,\Kelvin$.  N.B.\,the different
  temperature ranges are used in the colour indeces for the upper and
  lower panels. The temperature maps from the lower panels are adopted
  in the line profile calculation (see Section~{\ref{sub:depend-Tmax}}).
  The high density disc zone,
  indicate by the \emph{dashed lines} is excluded from our line
  profile calculations. See Section~\ref{sub:depend-Tmax} for
  explanations. The length scales are in the units of the stellar radius
  ($R_{*}$). }

\label{fig:temperature}

\end{figure*}

We now investigate whether the MHD outflows found in Model~B of the
previous section is a plausible solution for CTTSs, by computing line
profiles based on the simulation, and by checking the consistency
with the existing spectroscopic observations. The stellar parameters
adopted here are same as in Section~\ref{sub:mhd-setup}, and the
continuum sources as described in Section~\ref{sub:continum-sources}.
The density, velocity and temperature 
(but see Section~\ref{sub:depend-Tmax}) 
found in Model~B are mapped
on to the radiative transfer grid (e.g.~KU11),
and used as input variables in the profile calculation. We exclude
the high density disc accretion disc region (indicated by the dashed
line in Fig.~\ref{fig:temperature}), and concentrate on modelling
the line emission and absorption only from the accretion funnel and
the outflow regions. The disc region is excluded from the computations
because of the gas temperature near
the inner part of the disc, especially near the inner edge of the
disc, is too high ($T\sim10^{5}\,\Kelvin$) in the simulation result
from Model~B (Fig.~\ref{fig:temperature}). If the relatively
high temperature gas near the inner edge of the disc is included,
the line emission is dominated from this region, and the line strengths
are unrealistically strong. The disc temperature at larger radii ($r>20\, R_{*}$)
are low 1,500--2,000~K, and we do not expect any significant contribution
to the atomic hydrogen and helium emission. To mimic the opaqueness
of the accretion disc, we simply place a geometrically thin but optically
thick disc on the equatorial plane with the inner radius corresponding
to the inner radius of the high density MHD disc region (the vertical
dashed line in Fig\@.~\ref{fig:temperature}) which is removed from
the radiative transfer calculation. 

In the following, we examine the
dependency of the line profiles on our main model parameters: the
maximum gas temperature ($T_{\mathrm{max}}$), X-ray luminosity ($L_{\mathrm{X}}$)
and the temperature of the X-ray emitting gas ($T_{\mathrm{X}}$)
(cf.~Section~\ref{sub:cont-x-ray}). The model parameters are summarised
in Table~\ref{tab:Profile-Model-Summary}. To examine the dependency
of the line profiles on the model parameters, we mainly focus on modelling
the lines which often show a sign of wind (a blueshifted absorption
component), namely H$\alpha$ and \ion{He}{i}~$\lambda10830$. 
  In the following two sections (Sections~\ref{sub:depend-Tmax} and
  \ref{sub:depend-X-ray}), example line profiles are calculated at
  $i=30^{\circ}$ because \ion{He}{i}~$\lambda10830$ profile exhibits
  both the blueshifted wind absorption and the redshifted
  magnetospheric absorption most clearly at this inclination
  angle. However, readers should be aware that the probabilities of
  observing an object with $i<30^{\circ}$ and $i>30^{\circ}$ are 0.134
  and 0.866, respectively.  
Example calculations of additional hydrogen and helium lines, and their
dependency on the inclination angle ($i$) will be also presented in
Section~\ref{sub:Depend-inc}. 


%
\begin{table}

\caption{Summary of line profile model parameters}

\label{tab:Profile-Model-Summary}

\begin{center}

\begin{tabular}{rcccc}
\hline

 & 
$T_{\mathrm{max}}$
 & 
$L_{\mathrm{X}}$
 & 
$T_{\mathrm{X}}$
 &
$\gammaHeI$
\tabularnewline

Model ID
 & 
{(}$10^{3}\,\Kelvin${)}
 & 
{(}$\mathrm{10^{30}\, erg\, s^{-1}}${)}
 & 
{(}$10^{6}\,\Kelvin${)}
 &
{(}$10^{-4} s^{-1}${)}
\tabularnewline
\hline 

B1
 & 
$8$
 & 
$4$
 & 
$2$
 &
$1.2$
\tabularnewline

B2
 & 
$10$
 & 
$4$
 & 
$2$
 &
$1.2$
\tabularnewline

B3
 & 
$14$
 & 
$4$
 & 
$2$
 &
$1.2$
\tabularnewline

B4
 & 
$10$
 & 
$0.4$
 & 
$2$
 &
$0.12$
\tabularnewline

B5
 & 
$10$
 & 
$20$
 & 
$2$
 &
$6.0$
\tabularnewline

B6
 & 
$10$
 & 
$4$
 & 
$5$
 &
$0.2$
\tabularnewline

B7
 & 
$10$
 & 
$4$
 & 
$20$
 &
$0.012$
\tabularnewline
\hline
\end{tabular}

\end{center}

\end{table}



\subsection{Dependency on the maximum gas temperature}

\label{sub:depend-Tmax}

Our initial calculations of the line profiles based on the MHD
simulation Model~B, with unmodified temperature structure from the
simulations,  
showed that the line strengths (e.g.~H$\alpha$) are unrealistically
too strong when compared with observations
(e.g.~\citealt{reipurth:1996}).  This is mainly due to the emission
from the relatively high temperature gas ($\sim10^{5}\,\Kelvin$) near
the interface between the relatively high density conical-shell wind
region and the low density polar wind region
(Fig.~\ref{fig:temperature}).  The original temperature of the flows
in the outflows are, in general, higher than the typical wind
temperatures used in our previous line profile models
($\sim10^{4}\,\Kelvin$, KU11) which showed reasonable agreement with
the observed CTTSs.  A recent local excitation study for line
opacities of CTTSs by \citet{Kwan:2011} has also found a similar
temperature range.
  When the gas from the inner part of the disc is
  ejected to the wind, it cools down due to the adiabatic
  expansion, which is treated properly in the MHD
  simulations. However, the additional radiative cooling mechanism,
  which is not included in the simulation, may be required to cool
  down the gas even further. In general, the temperatures from the MHD
  simulations could be quite uncertain. We therefore modify the
  original gas temperature of the MHD simulations, in order to match
  observed line profiles, and treat the temperature as though a free
  parameter in the line profile calculations.
For this reason, we limit the
maximum temperature ($T_{\mathrm{max}}$) of the gas from the MHD
simulation to be $\sim10^{4}\,\Kelvin$ when it is applied to the
radiative transfer models. 

  The lower panels in Fig.~\ref{fig:temperature} show
  the temperature maps of MHD Model~B when the maximum allowed
  temperature $T_{\mathrm{max}}=10^{4}\,\Kelvin$ is applied.  The
  figures show that $T_{\mathrm{max}}$ not only affects the low
  density and high temperature polar wind regions, but also affect the
  conical-shell wind region. The relatively high density conical-shell
  wind region (including its base), is essentially isothermal (with
  $T=T_{\mathrm{max}}=10^{4}\,\Kelvin$) in this case.  On the other
  hand, the original temperature of the accretion funnel from the MHD
  simulation (the upper panels in Fig.~\ref{fig:temperature}), is
  relatively low compared to the wind temperature and is mostly less
  than $\sim10^{4}\,\Kelvin$. However, the temperatures in the funnel
  flow near the stellar surface and near the accretion disk are
  slightly higher than $10^{4}\,\Kelvin$ (in the original temperature
  from the MHD simulation); hence, they are affected by the imposition
  of $T_{\mathrm{max}}$.
In the following, we examine the effect of
changing $T_{\mathrm{max}}$ on the computed line profiles.

Fig.~\ref{fig:Tx-test} shows the model line profiles for H$\alpha$ and
\ion{He}{i}~$\lambda10830$ computed for
$T_{\mathrm{max}}=$~$8.0\times10^{3}$, $1.0\times10^{4}$ and
$1.4\times10^{4}\,\Kelvin$ (Models~B1, B2 and B3 in
Table~\ref{tab:Profile-Model-Summary}). The inclination angle, the
X-ray luminosity and temperature are fixed at $i=30^{\circ}$,
$L_{\mathrm{X}}=4\times10^{30}\,\mathrm{erg\, s^{-1}}$ and
$T_{\mathrm{X}}=2\times10^{6}\,\Kelvin$, respectively. The strengths
of the emission lines are comparable to those found in observations
(e.g.~\citealt{reipurth:1996}; \citealt{edwards:1994};
\citealt{Muzerolle:1998a}; \citealt{alencar:2000}; ED06) in this
temperature range. 

The line profile morphology of the \ion{He}{i}~$\lambda10830$ models
are also similar to some of the observed line profiles found in
ED06.  In particular, they are similar to those
which exhibit a narrow blueshifted absorption component that often
extends to $\sim200\,\kmps$. The relatively narrow blueshifted
absorption components, centred around $v\approx-120\,\kmps$, seen the
models are caused by the outflowing gas in the conical-shell wind, and the
redshifted absorption component, which shows the minimum flux at
$v\sim100\,\kmps$ and extends to $v\sim250\,\kmps$ is caused by the
inflowing gas that is accreting on to the stellar surface through the
accreting funnel. The figure shows that the emission strength and the
absorption components (in both red and blue wing) of
\ion{He}{i}~$\lambda10830$ are rather insensitive to the gas
temperatures in this range. This is because of the relatively large
ionization potential (24.6~eV) of \ion{He}{i}, and the plasma
temperature is not quite high enough for significant collisional
excitations. The main cause of the excitation for \ion{He}{i} in these
models are due to photoionizations e.g.\,by the X-ray flux (c.f.~KU11).

\begin{figure}

\includegraphics[clip,width=0.48\textwidth]{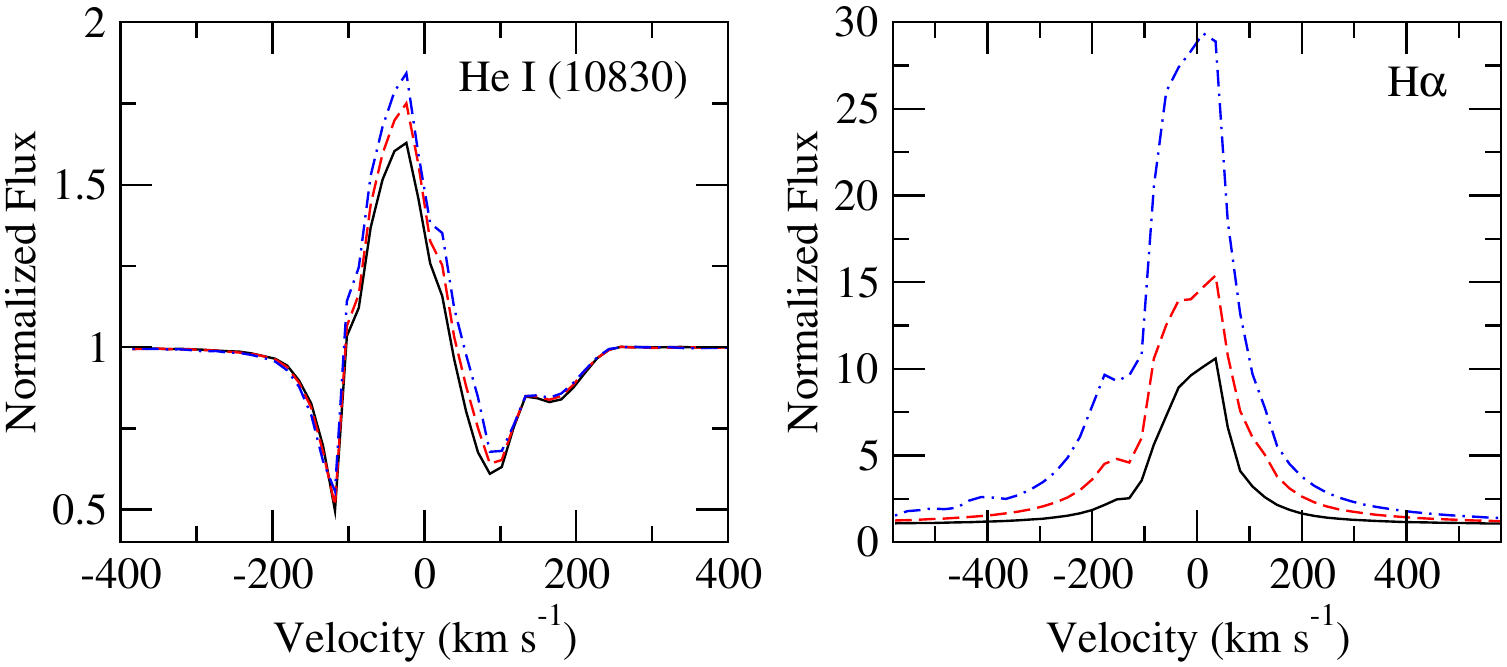}

\caption{Dependency of \ion{He}{i}~$\lambda$10830 (\emph{left}) and
  H$\alpha$ (\emph{right}) on the maximum gas temperature allowed
  ($T_{\mathrm{max}}$) in the profile calculations. MHD Model~B is
  used for all the models shown here. The profiles are computed at the
  fixed inclination angle $i=30^{o}$ and fixed X-ray luminosity and
  temperature, which are
  $L_{\mathrm{X}}=4\times10^{30}\,\mathrm{erg\,s^{-1}}$ and
  $T_{\mathrm{X}}=2\times10^{6}\,\Kelvin$, respectively. The models
  shown here use $T_{\mathrm{max}}=8.0\times10^{3}$ (\emph{solid}),
  $1.0\times10^{4}$ (\emph{dash}), and $1.4\times10^{4}\,\Kelvin$
  (\emph{dash-dot}), which corresponds to Models~B1, B2 and B3 in
  Table~\ref{tab:Profile-Model-Summary}, respectively.  }

\label{fig:Tmax-test}

\end{figure}

On the other hand, the emission strengths of the model H$\alpha$ is
sensitive to $T_{\mathrm{max}}$ value. The line emission becomes
stronger as $T_{\mathrm{max}}$ increases. Relatively weak but clearly
visible absorption components are seen in the blue wing of the line
($v\approx-120\,\kmps$). This, again, is caused by outflowing in the
conical-shell wind. Unlike \ion{He}{i}~$\lambda10830$, the redshifted
absorption component that goes below the continuum level is absent
from the H$\alpha$ profiles. This is consistent with observations
(e.g.~\citealt{reipurth:1996}) which find that H$\alpha$ profiles of
CTTSs rarely show redshifted absorption below the continuum. For
$T_{\mathrm{max}}=1.4\times10^{4}\,\Kelvin$, the line centre flux
reaches $\sim30$ (normalised to the continuum) which is close to the
maximum value found in the observations. The peak flux of the
H$\alpha$ profiles computed with
$T_{\mathrm{max}}=1.0\times10^{4}\,\Kelvin$ is $\sim15$ (normalised to
the continuum) which is similar to typical values found in
observations. 
  In summary, based on the line strengths of H$\alpha$ in
  Fig.~\ref{fig:Tmax-test} in comparison with those of typically
  observed H$\alpha$, the gas temperature of the accretion funnel and
  the conical-shell wind in this particular MHD simulation should be
  less than $\sim 1.4 \times 10^{4}\,\Kelvin$.
In the rest of this paper, we adopt
$T_{\mathrm{max}}=1.0\times10^{4}\,\Kelvin$ in our line profile
calculations.

Here, we have adopted a rather simplistic method of the gas
temperature assignment, with the introduction of $T_{\mathrm{max}}$
parameter. However, this leads to a large reduction of temperatures
(from $\sim 10^{6}\,\Kelvin$ to $\sim 10^{4}\,\Kelvin$) in the
low-density polar wind region where the gas temperature may be less
affected by the radiative cooling. The limiting the gas temperature to
$\sim 10^{4}\,\Kelvin$ in the polar wind region would be reasonable
when the wind density there is very low; hence, the omission of the
opacity in the polar wind region does not affect the line profiles. On
the other hand, if the polar wind density becomes relatively high, the
wind opacity may become high enough to affect the line profile;
therefore, the temperature of the polar wind should be treated more
properly.

\subsection{Dependency on X-ray luminosity and temperature}

\label{sub:depend-X-ray}

Next, we examine the dependency of the model profiles on the X-ray
luminosity ($L_{\mathrm{X}}$) and the temperature of the X-ray
emitting gas ($T_{\mathrm{X}}$). As in Section~\ref{sub:cont-x-ray},
our models assume that the X-ray emitting plasma is radiating
thermally (as a blackbody) with a single temperature
$T_{\mathrm{X}}$. The thermal X-ray radiation flux is normalised with
the total X-ray luminosity
$L_{\mathrm{X}}\left(0.1-10\,\mathrm{keV}\right)$. The photoionization
process by high energy photons are expected to be important for the
formation of the wind sensitive line, \ion{He}{i}~$\lambda10830$
(c.f.~\citealt{Kwan:2011}; KU11).

\begin{figure}

\begin{center}

\includegraphics[clip,width=0.48\textwidth]{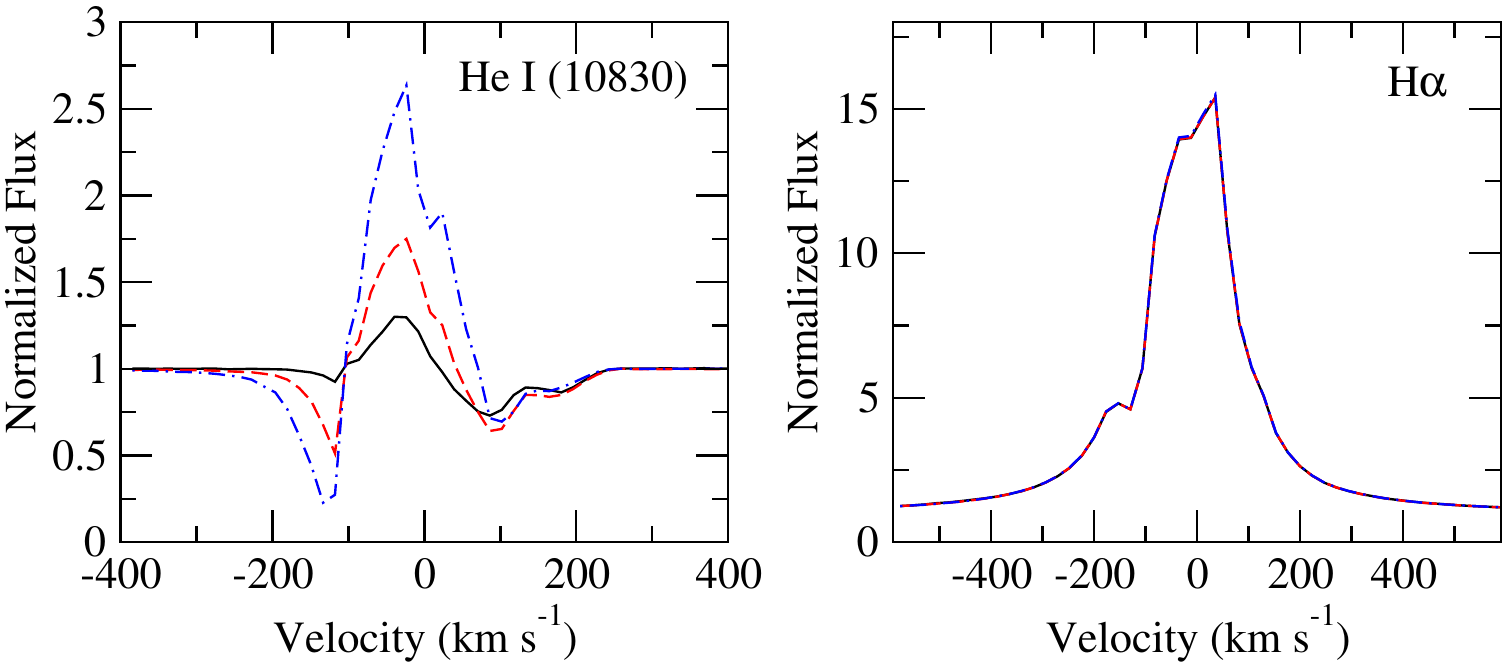}

\end{center}

\caption{Dependency of \ion{He}{i}~$\lambda$10830 and H$\alpha$
  profiles on the X-ray luminosity $L_{\mathrm{X}}$. The MHD
  simulation, Model~B, is used for all the models shown here. The
  profiles are computed at the fixed inclination angle $i=30^{o}$ and
  fixed X-ray emitting plasma temperature
  $T_{\mathrm{X}}=2\times10^{6}\,\Kelvin$ (assumed isothermal). The
  values of $L_{\mathrm{X}}$ used are $4 \times 10^{29}$
  (\emph{solid}), $4 \times 10^{30}$ (\emph{dash}) and $2 \times
  10^{31}\,\mathrm{erg \, s^{-1}}$ (\emph{dash-dot}), which
  corresponds to Models~B4, B2 and B5 in
  Table~\ref{tab:Profile-Model-Summary}, respectively.  The X-ray
  luminosity between $4 \times 10^{29}$ and $4 \times
  10^{31}\,\mathrm{erg \, s^{-1}}$ produces reasonable line strengths
  in \ion{He}{i}~$\lambda$10830 when compared to observations
  (e.g.~ED06).  The model H$\alpha$ profiles are not
  affected by the change in the values of $L_{\mathrm{X}}$, i.e.\,
  three line profiles are almost identical to each other.}

\label{fig:Lx-test}

\end{figure}

Fig.~\ref{fig:Lx-test} shows the model \ion{He}{i}~$\lambda$10830
profiles computed for $L_{\mathrm{X}}=$$4\times10^{29}$ ,
$4\times10^{30}$ and $2\times10^{31}\,\mathrm{erg\,
  s^{-1}}$(Models~B4, B2 and B5 in
Table~\ref{tab:Profile-Model-Summary}), which are within the range of
observed values for CTTSs (e.g.~\citealt{Telleschi:2007};
\citealt{Gudel:2007}; \citealt{Gudel:2010}). The inclination angle and
the X-ray temperature are fixed at $i=30^{o}$ and
$T_{\mathrm{X}}=2\times10^{6}\,\Kelvin$, respectively.
The corresponding photoionization rates
of \ion{He}{i} from the ground state ($\gammaHeI$) at a typical
location in the wind ($r=5.6R\,_{*}$) for Models~B4, 
B2 and B5 are $1.2\times 10^{-5}$, $1.2\times 10^{-4}$ and $6.0\times
10^{-4}\,\mathrm{s^{-1}}$, respectively
(Table~\ref{tab:Profile-Model-Summary}).  Since the X-ray flux is
directly proportional to a value of $L_{\mathrm{X}}$, the
photoionization rate $\gammaHeI$ is
also proportional to $L_{\mathrm{X}}$ (see
Table~\ref{tab:Profile-Model-Summary}). The photoionization rates
here are similar to those found in \citet{Kwan:2011}, i.e.\,$\gammaHeI
\approx 10^{-5}$ -- $10^{-4}\,\mathrm{s^{-1}}$ in the local
excitation calculation of the plasma, in the context of the classical
T Tauri stellar wind.

The X-ray luminosity between $4\times10^{29}$ and
$2\times10^{31}\,\mathrm{erg\, s^{-1}}$ (or
  equivalently $\gammaHeI$ between $1.2\times 10^{-5}$ and $6.0\times
  10^{-4}\,\mathrm{s^{-1}}$) produces resonable line strengths in
\ion{He}{i}~$\lambda10830$ when compared to observations (e.g.~ED06),
and the line is very sensitive to this parameter. As in the previous
tests for $T_{\mathrm{max}}$ (Fig.~\ref{fig:Tmax-test}), the model
\ion{He}{i}~$\lambda10830$ profiles show both the redshifted
absorbtion casued by the accretion funnel and the blueshifted
absorption component casued by the conical-shell wind. The figure
shows that the line centre flux becomes larger as $L_{\mathrm{X}}$ 
(or equivalently $\gammaHeI$)
increases. Similarly, the depths of the absorption components on the
blue sides become deeper and wider as the value of $L_{\mathrm{X}}$ 
(or equivalently $\gammaHeI$)
increases. Interestingly, for the model with the largest
$L_{\mathrm{X}}$ ($2\times10^{31}\,\mathrm{erg\, s^{-1}}$), the blue
absorption component extends to $\sim-200\,\kmps$ which is larger than
a typical maximum speed observed in the conical-shell wind (see
Fig.~\ref{fig:MHD-Vel}).  Although the deepest and main part of the
blueshifted absorption ($\sim-120\,\kmps$) is casued by the relatively
high density conical-shell wind, the shallower but higher velocity
absorption is cased by the lower density and higher velocity
(Fig.~\ref{fig:MHD-Vel}) gas in the polar wind. For the largest
$L_{\mathrm{X}}$ model, there are enough high energy photons to
photoionize \ion{He}{i} in the polar wind, and the
\ion{He}{i}~$\lambda10830$ opaicty in the polar wind becomes
non-negligible.

The figure also shows the model H$\alpha$ profiles computed for the
same set of $L_{\mathrm{X}}$ values. Unlike
\ion{He}{i}~$\lambda10830$, H$\alpha$ is insentitive to the value of
$L_{\mathrm{X}}$, and the computed profiles for all three
$L_{\mathrm{X}}$ values are almost indentical to each other. The model
H$\alpha$ profile does not depend on the value of $L_{\mathrm{X}}$
mainly because the photoionization rates for `\ion{H}{i}' in the X-ray
energy range ($0.1-10\,\mathrm{keV}$) used here is small compared to
the collinsional rates in the relatively high density gas in the
funnel flow. A weak blueshifted absorption component, casued by the
conical-shell wind, is also seen at $v\sim-120\,\kmps$.  Unlike the
blueshifted absorption component seen in the
\ion{He}{i}~$\lambda10830$ model with the largest $L_{\mathrm{X}}$,
the wind absorption here does not extend to a high velocity
(e.g.~$\sim-200\,\kmps$); hence, the polar wind is not contributing to
the absorption in the H$\alpha$ model profiles here.

\begin{figure}

\includegraphics[clip,width=0.48\textwidth]{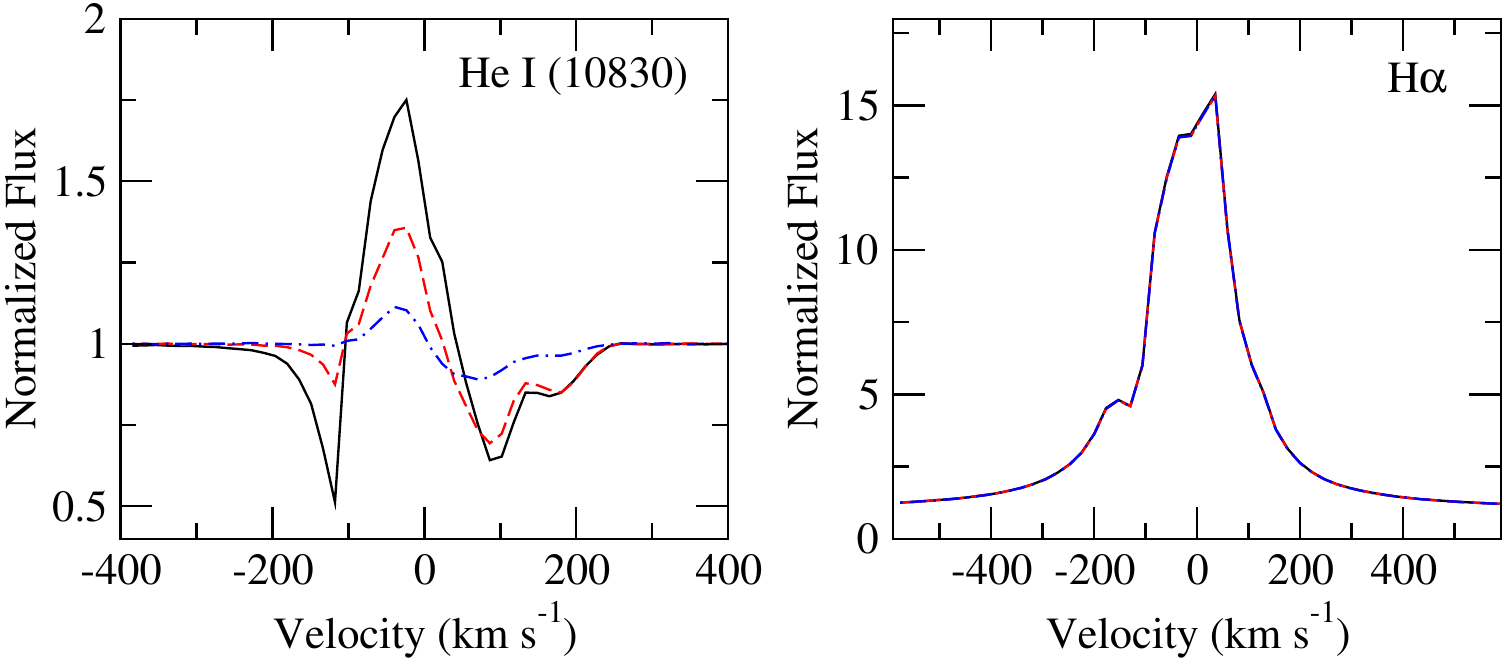}

\caption{Dependency of \ion{He}{i}~$\lambda$10830 and H$\alpha$
  profiles on the X-ray temperature $T_{\mathrm{X}}$. The MHD
  simulation, Model~B, is used for all the models shown here. The
  profiles are computed at the fixed inclination angle $i=30^{o}$ and
  with the fixed X-ray luminosity
  $L_{\mathrm{X}}=4\times10^{30}\,\mathrm{erg\,s^{-1}}$. The values of
  $T_{\mathrm{X}}$ used are $2\times10^{6}$ (\emph{solid}), $5\times
  10^{6}$ (\emph{dash}) and $2\times 10^{7}\,\Kelvin$
  (\emph{dash-dot}), which corresponds to Models~B2, B6 and B7 in
  Table~\ref{tab:Profile-Model-Summary}, respectively.  The X-ray
  temperature between $2 \times 10^{6}$ and $2 \times
  10^{7}\,\mathrm{erg \, s^{-1}}$ produces reasonable line strengths in
  \ion{He}{i}~$\lambda$10830 when compared to observations
  (e.g.~ED06). The model H$\alpha$ profiles are not
  affected by the change in the values of $T_{\mathrm{X}}$, i.e.\,
  three line profiles are almost identical to each other.}

\label{fig:Tx-test}

\end{figure}

We now examine the effect of changing the X-ray temperature
$T_{\mathrm{X}}$ on the profiles. Fig.~\ref{fig:Tx-test} shows the
model \ion{He}{i}~$\lambda$10830 profiles computed for
$T_{\mathrm{X}}=$$2\times10^{6}$, $5\times10^{6}$ and
$2\times10^{7}\,\Kelvin$ (Models~B2, B6 and B7, respectively in
Table~\ref{tab:MHD-Model-Summary}). The inclination angle and the
X-ray luminosity are fixed at $i=30^{\circ}$ and
$L_{\mathrm{X}}=4\times10^{30}\,\mathrm{erg\, s^{-1}}$ for all three
models. 
The corresponding photoionization rates
of \ion{He}{i} ($\gammaHeI$) at a typical
location in the wind ($r=5.6\,R_{*}$) for Models~B2, 
B6 and B7 are $1.2\times 10^{-4}$, $2.0\times 10^{-5}$ and $1.2\times
10^{-6}\,\mathrm{s^{-1}}$, respectively (see
Table~\ref{tab:Profile-Model-Summary}). 
For a fixed $L_{\mathrm{X}}$ value and in this temperature range,
the photoionization rate of \ion{He}{i} ($\gammaHeI$)
decreases as $T_{\mathrm{X}}$ increases because the peak of the blackbody
radiation curve shifts to a higher wavelength. Since the X-ray flux is
normalized to a fixed value of $L_{\mathrm{X}}$, this leads to 
a reduction of the softer X-ray flux when $T_{\mathrm{X}}$ is
increased. Consequently, the photoionization rate of \ion{He}{i}
decreases because its photoionization cross section is much larger 
towards the EUV frequency. 

The line strengths of the model \ion{He}{i}~$\lambda$10830
profiles computed with this range of $T_{\mathrm{X}}$ are comparable
similar to those found in the observations
(e.g.~ED06).  The line is sensitive to the change in
the value of $T_{\mathrm{X}}$ 
(or equivalently $\gammaHeI$).  
The flux at the line centre decreases
as $T_{\mathrm{X}}$ increases 
(or equivalently as $\gammaHeI$ decreases), 
and both blueshifted and redshifted
absorption components becomes weaker as $T_{\mathrm{X}}$
increases 
(or equivalently $\gammaHeI$ decreases). 
Similar to the earlier cases (Fig.~\ref{fig:Lx-test}), the model
H$\alpha$ profiles are~not sensitive to the change in $T_{\mathrm{X}}$
value, i.e.~the model profiles are almost identical to each
other. Again, the line is insensitive to the change in
$T_{\mathrm{X}}$ or more precisely to change in the X-ray flux because
the photoionization rates for \ion{H}{i} in the X-ray energy range
($0.1-10\,\mathrm{keV}$) is small compared to the collisional rates.

  As mentioned earlier (Section~\ref{sub:cont-x-ray}), the
  attenuation of the X-ray flux is neglected in our level population
  calculations; hence, it is implicitly assumed that the X-ray
  radiation is optically thin. 
  To check the validity of this assumption, we have calculated the
  X-ray optical depth ($\tau_{\mathrm{X}}$) at the photon energies
  $h\nu=0.1, 1.0$ and $10.0\,\mathrm{keV}$ along the line of sight to an
  observer located at the inclination $i=60^{\circ}$. We found   
  $\tau_{\mathrm{X}} \approx 90, 0.4$ and $ 0.01$ for $h\nu=0.1, 1.0$
  and $10.0\,\mathrm{keV}$, respectively.  This suggests that the X-ray
  flux should be almost completely attenuated for all photons with
  $h\nu<1.0$~keV. On the other hand, the X-ray flux at $h\nu>1.0$~keV are
  unaffected by the X-ray opacity.

\begin{figure}

\includegraphics[clip,width=0.48\textwidth]{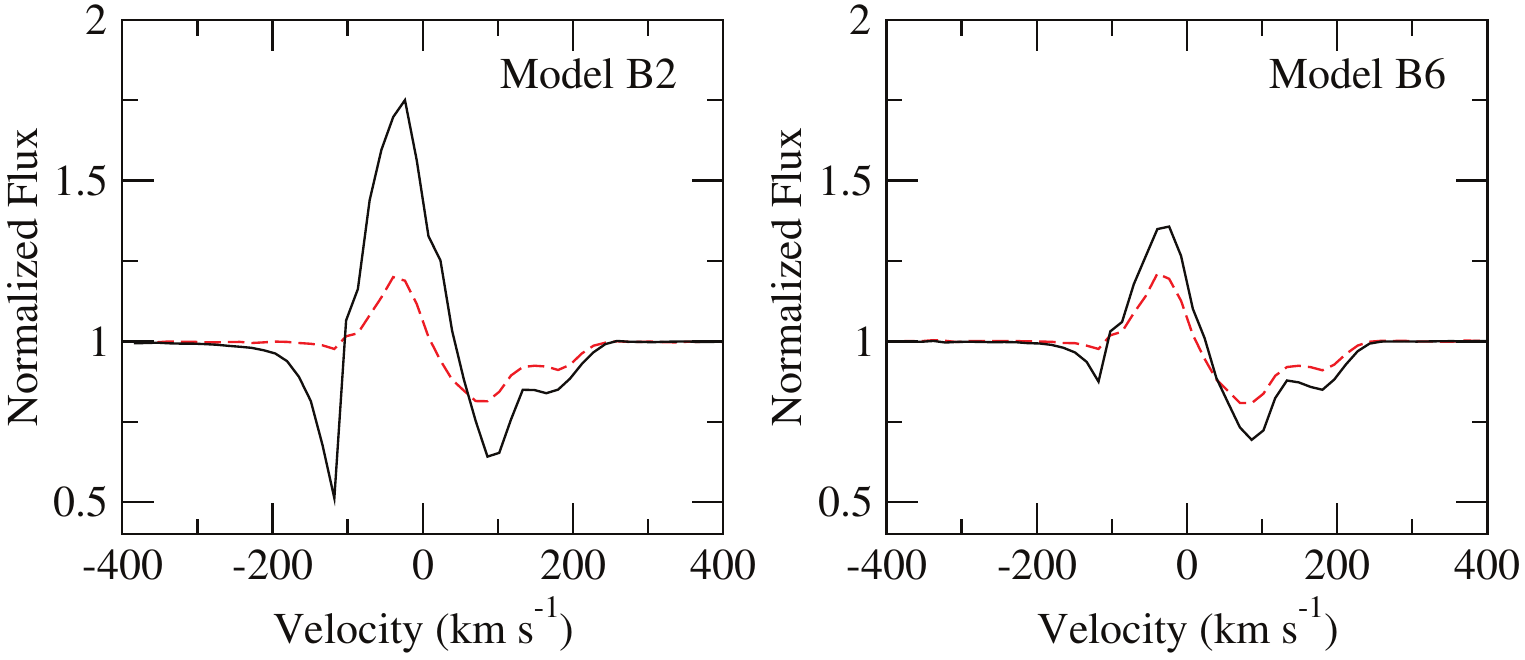}

\caption{
    Demonstration of the effect of the X-ray opacity on the formation
    of \ion{He}{i}~$\lambda$10830.  The \ion{He}{i}~$\lambda$10830
    line profiles from Models~B2
    ($T_{\mathrm{X}}=2\times10^{6}\,\Kelvin$) and B6
    ($T_{\mathrm{X}}=5\times10^{6}\,\Kelvin$) (\emph{solid}) are
    compared with the same models computed with the input X-ray
    continuum fluxes corrected for the attenuation by the continuum
    optical depth in X-ray (\emph{dash}).  Note that the original
    Models ~B2 and B6 have the same X-ray luminosity,
    $L_{\mathrm{X}}=4\times10^{30}\,\mathrm{erg\,s^{-1}}$
    (Table~\ref{tab:Profile-Model-Summary}). For a relatively low
    X-ray temperature ($T_{\mathrm{X}}=2\times10^{6}\,\Kelvin$)
    (\emph{left panel}), the reduction of the line emission and
    absorption strengths due to the attenuation of the X-ray flux is
    significant. The effect of the X-ray attenuation becomes
    smaller for the model with a higher X-ray temperature
    ($T_{\mathrm{X}}=5\times10^{6}\,\Kelvin$) (\emph{right panel}).
}

\label{fig:taux-test}

\end{figure}

  To assess the effect of the X-ray
  attenuation on the model profiles, we do the following: (1)~compute the
  continuum optical depth $\tau_{\mathrm{X}}$ at all frequency in
  X-ray, (2)~attenuate the X-ray flux by a factor of $e^{-\tau_{\mathrm{X}}}$
  at each X-ray frequency, (3)~recompute line profiles
  (including source function calculations) using the X-ray flux
  corrected for the attenuation due to X-ray opacity, and (4)~compare
  the new line profiles with the models computed without the
  attenuation of the X-ray flux. 
  For this test, we use \ion{He}{i}~$\lambda$10830 line only since
  H$\alpha$ profiles are rather unaffected by the X-ray flux as
  previously seen in Figs.~\ref{fig:Lx-test} and \ref{fig:Tx-test}.
  We have recomputed the \ion{He}{i}~$\lambda$10830 line profiles for
  Models~B2 and B6 (Table~\ref{tab:Profile-Model-Summary}) with the
  X-ray flux with the attenuation.  Note that these two models have 
  the X-ray luminosity
   $L_{\mathrm{X}}=4\times10^{30}\,\mathrm{erg\,s^{-1}}$ but have
  different X-ray temperatures, $T_{\mathrm{X}}=2\times10^{6}$ and 
  and $5\times10^{6}~\Kelvin$, respectively.  The results are shown in
  Fig.~\ref{fig:taux-test}.

The figure shows that the reduction of the line emission and
absorption strengths due to the attenuation of the X-ray flux is
significant, for a relatively low X-ray temperature (Model~B2:
$T_{\mathrm{X}}=2\times10^{6}\,\Kelvin$).  However, the effect of the
X-ray attenuation becomes notably smaller for the model with a higher
X-ray temperature (Model~B6: $T_{\mathrm{X}}=5\times10^{6}\,\Kelvin$).
The difference is mainly caused by the difference in the flux
distributions of the blackbody radiation curve for different X-ray
temperatures. The X-ray flux of the lower X-ray temperature
($T_{\mathrm{X}}=2\times10^{6}\,\Kelvin$) peaks at a lower energy
($h\nu \approx 0.5$~keV) than that of the higher temperature
($T_{\mathrm{X}}=5\times10^{6}\,\Kelvin$) which peaks at $h\nu \approx
1.2$~keV.  As mentioned earlier, the X-ray attenuation is important
mostly for the photon energies $h\nu<1.0$~keV. This indicates that a
larger fraction of X-ray flux is affected/attenuated for the blackbody
radiation curve with the smaller temperature
($T_{\mathrm{X}}=2\times10^{6}\,\Kelvin$) since its peak is located
below $1.0$~keV. Consequently, the photoionization rate of \ion{He}{i}
is less affected for the higher $T_{\mathrm{X}}$ model (Model~B6);
hence, the line profile for \ion{He}{i}~$\lambda$10830 is less
affected for the higher $T_{\mathrm{X}}$.

From this test, it is clear that the assumption of no X-ray
attenuation could significantly overestimate the line emission and
absorption strengths in \ion{He}{i}~$\lambda$10830. In this test, we
have considered the models with the X-ray flux attenuation which is
calculated from the opacity of the models with no X-ray attenuation;
hence, they are not completely self-consistent. This method may be
acceptable for obtaining rough estimates of the effect of the X-ray
attenuation on the line profiles.  However, to implement the X-ray
attenuation effect properly in our radiative transfer model (KU11),
the radiation field (including the X-ray radiation) and opacity must
be solved simultaneously. This requires an iteration between the
opacity and the radiation field calculations until they are mutually
consistent.  Unfortunately, our radiative transfer code (KU11) is not
designed for this type of iterations, and we have to assume that the
gas is optically thin for the underlying continuum source. This is a
limitation our radiative transfer model and is an important
caveat. Therefore, we warn readers that the emission and absorption
strengths of \ion{He}{i} lines presented in this work could be
overestimated significantly due to the assumption of no X-ray
attenuation in our model.

  However, in reality, the magnetospheric accretion funnel, which is a
  significant contributor to the X-ray optical depth, may not be
  axisymmetric, and it could occur in two or more funnel streams,
  as shown in 3D MHD simulations (e.g. \citealt{Romanova:2003}). This
  flow geometry may contain large gaps between the funnel flows, and it
  could allow a large fraction of the X-ray produced near the stellar
  surface to reach the conical-shell regions without a significant
  attenuation, and ionize \ion{He}{i} in the wind.


\subsection{Dependency on inclination angle and additional lines }

\label{sub:Depend-inc}

\begin{figure*}

  \includegraphics[clip,width=1\textwidth]{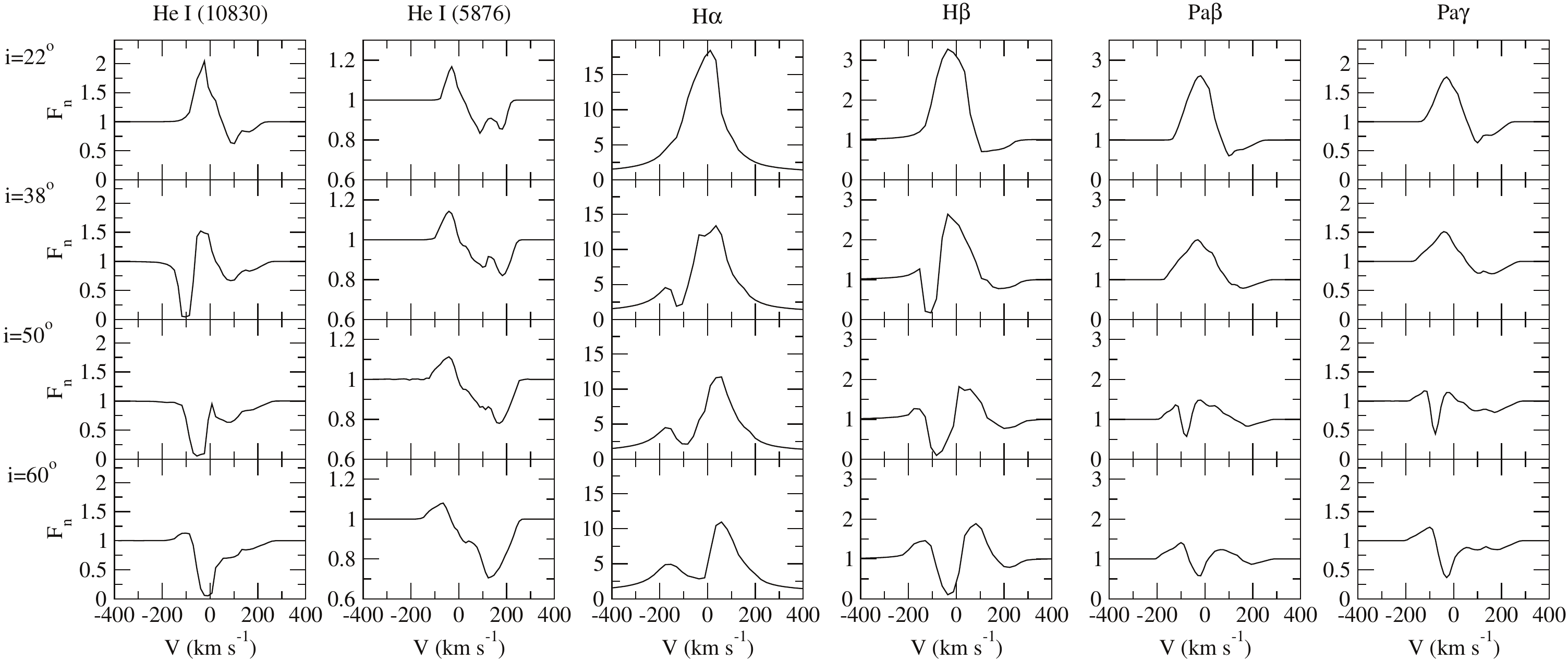}

  \caption{Summary of sample model hydrogen and helium line profiles
    computed at four different inclination angles: $i=22^{\circ}$, $38^{\circ}$,
    $50^{\circ}$ and $60^{\circ}$ 
    (equally spaced in $\cos i$ between $i=0^{\circ}$ and
   $65^{\circ}$)  
    from the top to bottom rows,
    respectively.  From the left to right columns, sets of
    \ion{He}{i}~$\lambda$10830, \ion{He}{i}~$\lambda$5876, H$\alpha$,
    H$\beta$, Pa$\beta$ and Pa$\gamma$ profiles are shown
    respectively. The X-ray luminosity
    $L_{\mathrm{X}}=4\times10^{30}\,\mathrm{erg\,s^{-1}}$ and
    temperature $T_{\mathrm{X}}=2\times10^{6}\,\Kelvin$ (Model~B2 in
    Table~\ref{tab:Profile-Model-Summary}) are used for all the models
    shown here.  The fluxes ($F_{\mathrm{n}}$) are normalised to the
    local continuum. }
  
  \label{fig:inc-test}

\end{figure*}

We now examine the dependency of the model profiles on the inclination
angles ($i$) and study the general characteristics by using the set of
model parameters used for Model~B2 (see
Table~$\ref{tab:Profile-Model-Summary}$), i.e.~
$T_{\mathrm{max}}=1.0\times10^{4}\,\Kelvin$,
$L_{\mathrm{X}}=4\times10^{30}\,\mathrm{erg\, s^{-1}}$ and
$T_{\mathrm{X}}=2\times10^{6}\,\Kelvin$. In addition to
\ion{He}{i}~$\lambda$10830 and H$\alpha$, samples of
\ion{He}{i}~$\lambda$5876, H$\beta$, Pa$\beta$ and Pa$\gamma$ line
profiles are also computed. 
Since the accretion disc region ($\theta \simgreat 65^{\circ}$ as shown
in Fig.~\ref{fig:temperature}) is excluded in the opacity and
emissivity calculations, the line profiles presented hereafter are
restricted to $i < 65^{\circ}$. This is done also to avoid the
complication that arises from the occultation of the line emission by
the extended disc which we do not treat self-consistently. 
We choose four different inclination
angles, equally spaced in $\cos i$ between  $ i = 0^{\circ}$ and
$65^{\circ}$, to present the profiles of equal probability of
occurrence between the inclination angle range, i.e.\, they are
approximately $i=22^{\circ}$, $38^{\circ}$, $50^{\circ}$ and
$60^{\circ}$. The results are summarized in Fig.~\ref{fig:inc-test}.  
In general, the
line strengths of the model profiles are comparable to those
found in observations (e.g.~\citealt{reipurth:1996};
\citealt{edwards:1994}; \citealt{Muzerolle:1998a};
\citealt{alencar:2000}; \citealt{folha:2001}; ED06), except for
\ion{He}{i}~$\lambda$5876 which are much weaker than a typical value
from observations (see e.g.~\citealt*{Beristain:2001}).  In the
following, the characteristics of each line are examined more
closely. 

\emph{\ion{He}{i}~$\lambda$10830.} The conical-shell wind geometry with the
half-opening angle $\sim35^{\circ}$ (Section~\ref{sub:MHD-Model-B})
does not favour the wind absorption at a low inclination
(e.g.~$i=22^{\circ}$).  
The blueshifted absorption, caused by the
conical-shell wind, starts to appear at $i\approx30^{\circ}$. The position
of the blueshifted absorption shifts from $v\sim120\,\kmps$ (at
$i\approx38^{\circ}$) towards the line centre as the inclination becomes
higher (up to $i\approx60^{\circ}$).  This shift is caused by the
change in the line-of-sight velocity of the conical-shell wind toward the
observer as the inclination changes, The redshifted absorption,
caused by the accretion funnel (c.f.~Fig.~\ref{fig:MHD01}), is present for
all $i$. The extent of the redshifted absorption changes between
$\sim200$ to $\sim250\,\kmps$ depending on $i$, 
which is similar to the flow speed of the gas in the accretion funnel
near the stellar surface (c.f.\,Section~\ref{sub:MHD-Model-B}).  
At higher inclination angles ($i \simgreat 50^{\circ}$), the line
emission almost disappears, and the profile is shaped by the wind and
funnel absorption features.


\emph{\ion{He}{i}~$\lambda$5876}. The line peak flux slightly decreases and
the redshifted absorption component becomes stronger as the inclination
increases. No blueshifted wind absorption component is seen in this
line for all $i$.  This is in agreement with the observation of
\citet{Beristain:2001}.  The redshifted absorption components, on the
other hand, are present for all $i$. However, the observations of this
line, from 31 CTTSs (\citealt{Beristain:2001}), show that the
redshifted absorption is rather rare (3 out 31). The line peak fluxes
in our models are between $\sim1.1$ and $\sim1.2$ (normalised to
continuum) which are much smaller than those found in the observations
(typically $>1.2$, e.g.~\citealt{alencar:2000};
\citealt{Beristain:2001}). This line is thought to arise from the post
shock region of the accretion funnel near the stellar surface
(e.g.~\citealt{Beristain:2001}; \citealt{Edwards:2003};
\citealt{Orlando:2010}; \citealt{Kwan:2011}). This indicates that a
higher temperature and density gas may be needed in order to match the
observations which would require more formal treatment of the shock
regions in the accretion funnels (e.g.~\citealt{Lamzin:1998};
\citealt{Koldoba:2008}; \citealt{Sacco:2008}; \citealt{Orlando:2010}).

\emph{H$\alpha$}. Unlike the helium lines,  no clear
redshifted absorption component is seen in this line. However, the
profile at the low inclination ($i=22^{\circ}$) shows a slight blue
asymmetry, i.e.~the flux distribution is slightly shifted toward
blue, which may be caused by a small amount of absorption in the red
wing although it is not seen as a discrete absorption component. At a
low inclination angle ($i\approx22^{\circ}$), no clear blueshifted
wind absorption component is seen in the model profiles. A wind
absorption becomes visible for $i \simgreat 38^{\circ}$, and its strength
increases as $i$ increases. In general, the models are in good
agreement with observations (e.g.~\citealt{reipurth:1996};
\citealt{edwards:1994}; \citealt{Muzerolle:1998a};
\citealt{alencar:2000}). The model profiles here belong to Type~I and
Type~III~B (which constitute about 60~per~cent of the 
observed H$\alpha$ profiles for CTTSs in \citealt{reipurth:1996}) according to
the morphological classification of \citet{reipurth:1996}.

\emph{H$\beta$}. The redshifted absorption component is present in the
profiles at all $i$.  The extent of the redshift
  absorption changes between $\sim200$ to $\sim250\,\kmps$ depending on $i$. 
While no blueshifted absorption component is seen
at the lower inclination angles (e.g.~$i=22^{\circ}$), a
strong blueshifted wind absorption component is present 
at higher inclinations ($i>38^{\circ}$). 
The position of the wind absorption component shifts
toward the line centre as $i$ further increases. At $i=60^{\circ}$,
the wind absorption component is located near $v=0\,\kmps$, and
the resulting line has a double peaked profile shape.  At a very
high inclination angle, the conical-shell wind is moving almost
perpendicular to the line of sight to an observer; hence, the wind
absorption is located near the line centre. The double-peak profile
resembles that from a rotating disc; however, the H$\beta$ emission
here mainly arises from the magnetospheric accretion funnel whose rotational
velocity is rather slow ($\sim 20\,\kmps$).  The extent of the
emission in the blue wing becomes larger as $i$ increases. This is
consistent with the trends found in the radiative transfer models of
\citet{hartmann:1994} and \citet{muzerolle:2001} who modelled the line
emission from the axisymmetric magnetospheric accretion funnel flows.
The line profile shapes and the strengths found here are comparable to
those found in the observations (e.g.~\citealt{edwards:1994};
\citealt{alencar:2000}). However, the extent of the blue wing in the
model profiles ($\sim -130\,\kmps$ to $\sim -200\,\kmps$) tends to be
notably smaller than a typical value found in observations ($\sim
300\,\kmps$).

\emph{Pa$\beta$}. The inclination angle dependency of the line profiles are
somewhat similar to that of H$\beta$ and Pa$\gamma$. The redshifted
absorption component is visible at all $i$, and its depth slightly
decreases as $i$ increases.  The extent of the line emission in the
blue wing increases from $\sim-130\,\kmps$ to $\sim-200\,\kmps$ as $i$
increases from $22^{\circ}$ to $60^{\circ}$. At $i=22^{\circ}$ and
$38^{\circ}$, the models show a classic inverse P-Cygin profile, which
is classified as Type~IV~R profiles according to \citet{reipurth:1996}
and \citet{folha:2001}. This consists of about 34~per~cent (13 out of
37) of the objects observed by \citet{folha:2001}.  The blueshifted
wind absorptions are seen at the higher inclinations ($i=50^{\circ}$
and $60^{\circ}$), and their positions shifts towards the line centre
as $i$ increases. Similar to the H$\beta$ case, the wind absorption
component is located near $v\approx0\,\kmps$ for the model
with $i=60^{\circ}$, and the line exhibits a double peaked
shape. On contrary, the blueshifted wind component is extremely rare
(1 out of 37 objects) in the observed Pa$\beta$ profiles of
\citet{folha:2001}. This indicates that the Pa$\beta$ line opacity in
the conical-shell wind is too high possibly because the wind density and/or
temperature are too high in our model.

\emph{Pa$\gamma$.} The dependency on $i$ is very similar to that of
Pa$\beta$ and H$\beta$. The line shows the redshifted absorption
component at all $i$.  The frequency of the redshifted absorption
component is much lower (24~per~cent) in the observation of ED06. A
relatively strong blueshifted wind absorption appears in the
higher inclination angles, i.e.~$i>50^{\circ}$. The peak fluxes and
the profiles shapes of the models are similar to those found in the
observations of 38 CTTSs obtained by ED06, except for
the blueshifted absorption feature in the models with $i>50^{\circ}$.
No wind absorption component is  seen in the observed
Pa$\gamma$ profiles of ED06. As in the Pa$\beta$ models, this indicates
that the line opacity in the conical-shell wind is too high possibly because
the wind density and/or temperature are too high in our model. The
extent of the blue wing emission increases from 
$\sim-130\,\kmps$ to $\sim-200\,\kmps$ as $i$ increases from
$22^{\circ}$ to $60^{\circ}$. This behaviour, again, is consistent
with the line profile models of \citet{hartmann:1994} and
\citet{muzerolle:2001}.

\section{Discussion}

\label{sec:discussion}

\subsection{Contribution to line emissions from the conical-shell wind}

\label{sub:emiss-loc-test}

\begin{figure}

\begin{center}
  \includegraphics[clip,width=0.48\textwidth]{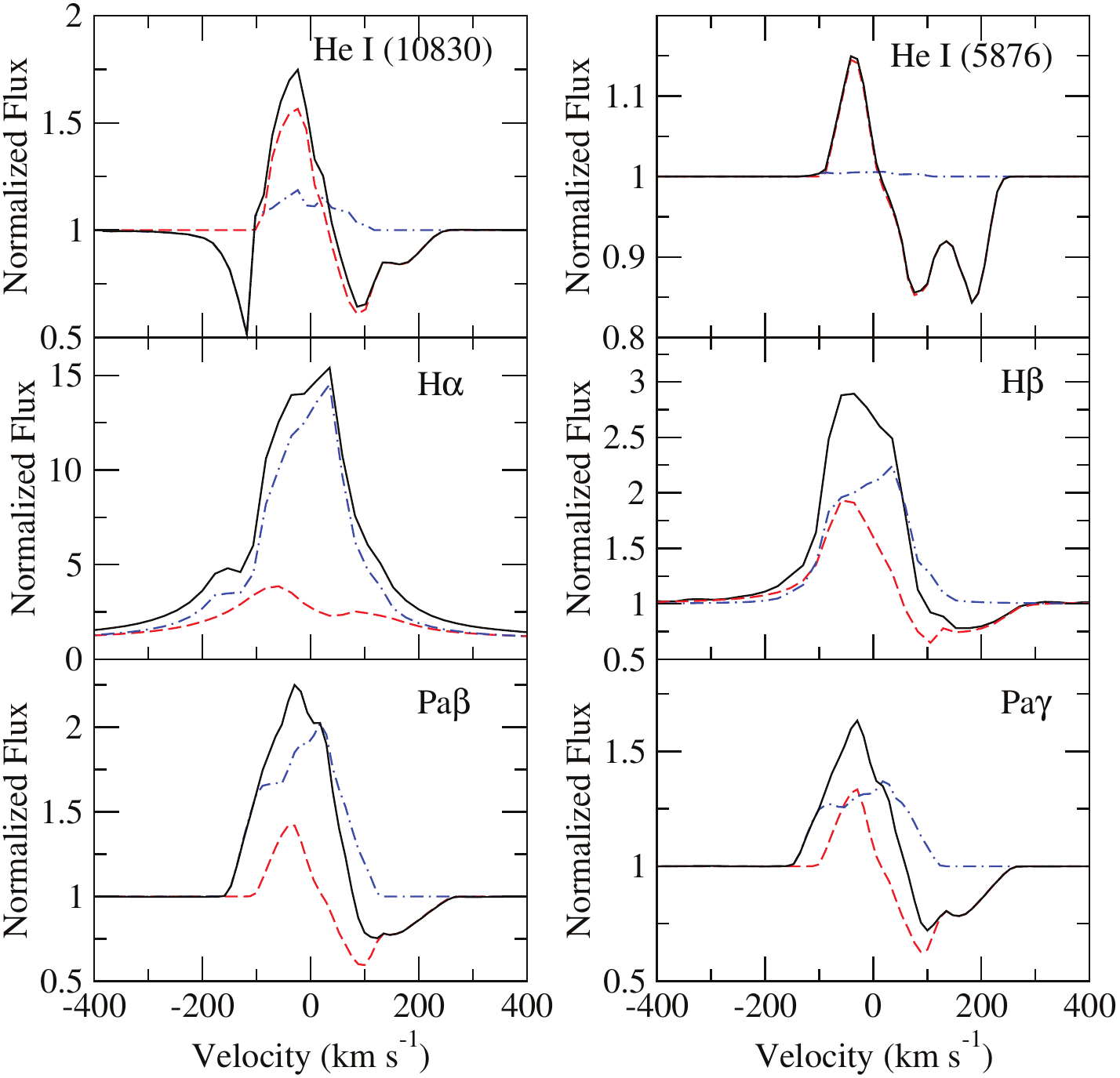}
\end{center}

\caption{
   Comparison of the model lines profiles (\ion{He}{i}~$\lambda$10830,
   \ion{He}{i}~$\lambda$5876, H$\alpha$, H$\beta$, Pa$\beta$ and
   Pa$\gamma$ ) computed with different flow elements in the MHD
   simulation (Model~B).  For each line, the profiles are computed
   with (1)~the accretion funnel part only 
   (\emph{dash}), (2)~the conical-shell wind part only
   (\emph{dash-dot}), and (3)~all the flow components -- the accretion
   funnel and the wind included (\emph{solid}). The physical
   parameters used here are
   the same as in Model~B2 (Table~\ref{tab:Profile-Model-Summary}),
   and the inclination angle is $i=30^{\circ}$.  Note that the flux
   contribution from the conical-shell wind in the hydrogen lines is sensitive
   to the assumed wind temperature which is essentially controlled by
   the model parameter $T_{\mathrm{max}}$
   (see~Section~\ref{sub:depend-Tmax}). The contribution from the wind
   would decrease significantly when $T_{\mathrm{max}}$ is decreased. 
}

\label{fig:emiss-loc-test}

\end{figure}

The model line profiles presented so far are the emergent profiles
which an observer at a given viewing angle would observe. This
emission consists of the emission from the inflowing gas (the
accretion funnel), the outflowing gas (the wind), and the continuum
emission.  Although the red and blueshifted absorption components in
the model line profiles could provide us some information on the
physical conditions and locations of the line `absorption', the
locations for the line `emission' could not be easily interpreted from
the total emergent line profiles. For this reason, we now examine the
relative contributions of the accretion funnel and the conical-shell
wind to the emergent profiles by comparing the line profiles computed
with (1) the accretion funnel part only, (2) the conical-shell wind
only, and (3) the combination of (1) and (2).  The physical parameters
used for this demonstration are the same as in Model~B2 (see
Table~\ref{tab:Profile-Model-Summary}). All the line profiles here are
computed at the inclination angle $i=30^{\circ}$.  The
results are placed in Fig.~\ref{fig:emiss-loc-test}.

The difference in the continuum levels for
each line profile is very small, i.e. the continuum emission from
the wind and the accretion funnel is negligible compared to the
stellar continuum; therefore, the line profiles presented in
Fig.~\ref{fig:emiss-loc-test} use the normalized flux for the vertical
scales.  The figure shows that, with this particular set of model
parameters, the emission in \ion{He}{i}~$\lambda$10830 mainly arises
from the accretion funnel (where the photoionization radiation is very
strong); however, the emission from the wind is non-negligible, and
contributing to the line centre flux. On the other hand,
the wind emission in \ion{He}{i}~$\lambda$5876 is negligibly small,
and the line emission is essentially from the accretion funnel. 
Unlike the helium lines, the wind emission in the hydrogen lines
 is comparable to that of the accretion funnel for H$\beta$ and
 Pa$\gamma$, and is significantly dominating that of the accretion
 funnel for H$\alpha$ and Pa$\beta$. However, the readers must be
 aware that the wind emission in hydrogen lines are very sensitive to 
the assumed wind temperature which is essentially controlled by our
model parameter $T_{\mathrm{max}}$, as demonstrated in
Section~\ref{sub:depend-Tmax} for H$\alpha$. The wind temperature in
these models is approximately isothermal with $T \approx
T_{\mathrm{max}}=10^{4}\,\Kelvin$; however, if a slightly lower wind
temperature, e.g.~$ T_{\mathrm{max}}=8\times 10^{3}\,\Kelvin$ is 
 used, the contribution from the wind emission will decrease
 significantly.

Despite the uncertainty in the wind temperature, it is interesting to
note that the extent of the blue wing in some hydrogen lines,
especially Pa$\beta$ and Pa$\gamma$, becomes notably ($\sim
50\,\kmps$) larger due to the wind emission. However, the amount of
the blue wing extention would depend on the relative flow speed of the
gas in the accretion funnel and the conical-shell wind. 
Since the wind emission in the hydrogen lines presented here is significant, the final
emergent line profiles (computed with both wind and accretion
funnel) appear broader than those from the accretion funnel or
magnetosphere alone. However, this also depends on the assumed wind
temperature.

\subsection{Comparisons of \ion{He}{i}~$\lambda$10830 profiles with observations}

\label{sub:compare-obs}

The sensitivity of \ion{He}{i}~$\lambda10830$ to the innermost winds
of CTTSs and its usefulness for probing the physical conditions of the
winds have been demonstrated by e.g.~ED06, \citet{Kwan:2007},
\citet{Kwan:2011} and KU11. Here, we examine whether the conical-shell wind
model (Model~B) can account for the types of
\ion{He}{i}~$\lambda10830$ profiles seen in the observations.  For
this purpose, we have run a small set of line profile models for a
various combinations of $L_{\mathrm{X}}$, $T_{\mathrm{X}}$ and $i$
while keeping other parameters fixed as in Section~\ref{sub:Depend-inc}. We
then examined if there is a resemblance between any of our model
profiles to a set of \ion{He}{i}~$\lambda10830$ line observations in
ED06. In the following, we present a several example cases in which
the model profile morphology is similar to that of observations.

\begin{figure}

\begin{center}
  \includegraphics[clip,width=0.48\textwidth]{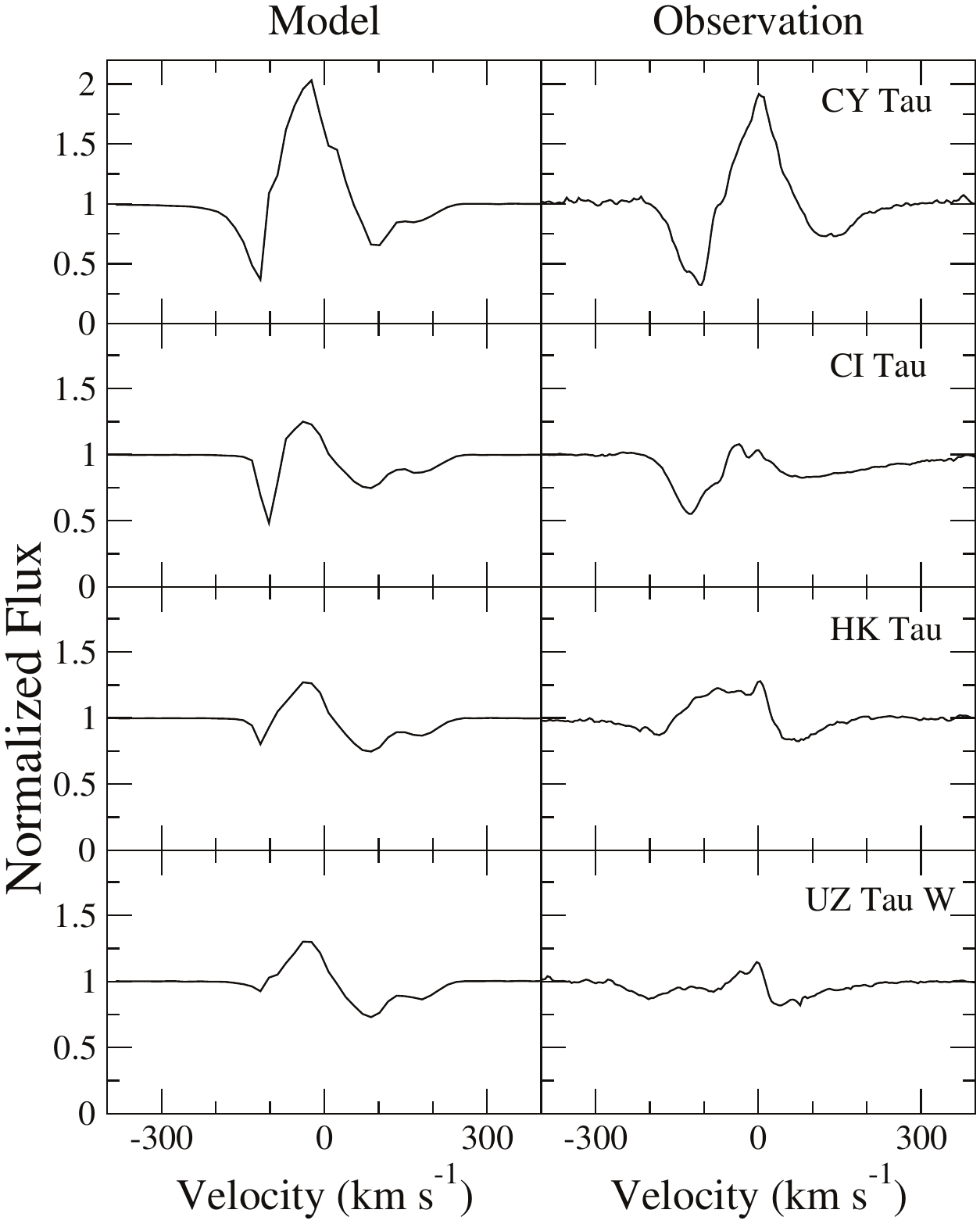}
\end{center}

\caption{Qualitative comparison of the model
  \ion{He}{i}~$\lambda$10830 profiles with the observation of
  ED06. Example model profiles (\emph{left column}) are selected based
  on their morphological similarities 
  with the observations (\emph{right column}).  For the
  observed profiles, the names of the objects are indicated on the
  upper-right corner of each panel. The corresponding model parameters
  used are summarised in Table~\ref{tab:compare-obs}. The
  \ion{He}{i}~$\lambda$10830 profiles predicted from the conical-shell wind
  model broadly agree with the subset of the observations, but the
  model does not show a deep and wide blueshifted (P-Cygni like) profiles (not
  shown here) found in some CTTSs (see~ED06).  }

\label{fig:compare-obs}

\end{figure}

Fig.~\ref{fig:compare-obs} shows qualitative comparisons of the model
\ion{He}{i}~$\lambda$10830 profiles with the observation of ED06. The
corresponding model parameters used are summarised in
Table~\ref{tab:compare-obs}.  The models shown here are not the strict
fits to the observations, but rather simple comparisons of the profile
morphology since we did not adjust the underlying stellar parameters
(stellar masses, radii, luminosity and so on) which could be
tailored to individual objects.
The figure shows that the line profiles predicted from the
conical-shell wind model (Model~B) broadly agree with the observations. 
The extent of the redshifted absorption, formed in the accretion
funnel flow (cf.~Fig.~\ref{fig:MHD01}), in the model line profiles is
$\sim 250\,\kmps$, which is comparable to those in the observed
profiles ($\sim 200$ to $\sim 300\,\kmps$).  The conical-shell wind produces a
relatively narrow blueshifted absorption component in the profile in
various degrees depending on the model parameters, $L_{\mathrm{X}}$, $T_{\mathrm{X}}$
and $i$. As we demonstrated in the previous section
(Figs.~\ref{fig:Lx-test}, \ref{fig:Tx-test} and \ref{fig:inc-test} in
Section~\ref{sec:results-profiles}), the wind absorption depths and
the emission strengths are sensitive to these parameters.

A relatively strong line centre emission (as seen in the case for
CY~Tau) can be obtained with a relatively high X-ray luminosity
($L_{\mathrm{X}}=8\times10^{30}\,\mathrm{erg\,s^{-1}}$) and a
relatively low inclination angle ($i=30^{\circ}$).  On the hand, a
relatively weak line centre emission and wind absorption (as seen in
CI~Tau, UZ~Tau~W and HK~Tau) can be obtained with a relatively low
X-ray
luminosity($L_{\mathrm{X}}=0.4$--$2\times10^{30}\,\mathrm{erg\,s^{-1}}$)
and a relatively low inclination angle ($i=30^{\circ}$--$36^{\circ}$).
A notable difference between the models and observation is in the
extent of the blueshifted absorption component in UZ~Tau~W.  In both
cases, the observations suggest a presence of high velocity outflows
(up to $\sim 300\,\kmps$) which cannot be explained by the
conical-shell wind in our MHD model because its maximum speed is only
$\sim 125\,\kmps$ (Fig.~\ref{fig:MHD-Vel}). The high velocity
component seen in the observations are possibly attributed to a
bipolar `stellar wind' which are not implemented in our model.

As demonstrated above, the conical-shell wind model (Model~B) are
capable of reproducing the relatively narrow and low-velocity
($v\approx -130$ to $-200\,\kmps$) blueshifted absorption component in
\ion{He}{i}~$\lambda$10830 profiles. However, the model does not show
the profiles with a deep and wide blueshifted wind absorption
component that extends to $v\approx -300$ to $-400\,\kmps$, which are
observed in about 40~per~cent of the samples in ED09.  These P-Cygni
like line profiles are likely caused by stellar winds that emerge in
the polar directions of CTTSs, as suggested by ED06,
\citet{Kwan:2007}, \citet{Kwan:2011} and KU11. The inability of our
model for reproducing the P-Cygni like profile was expected since our
MHD model does not include a stellar wind.

%
\begin{table}

\caption{Summary of models used for comparison of
  \ion{He}{i}~$\lambda$10830 with observations in
  Fig.~\ref{fig:compare-obs}.}

\label{tab:compare-obs}

\begin{center}

\begin{tabular}{rccl}
\hline 

Object & $L_{\mathrm{X}}$ & $T_{\mathrm{X}}$        &  inclination \tabularnewline
       & {(}$\mathrm{10^{30}\, erg\, s^{-1}}${)}&{(}$10^{6}\,\Kelvin${)}  & $\cdots$  \tabularnewline
\hline 
CY Tau   & $8$   &  $2$ &  $30^{\circ}$  \tabularnewline
CI Tau   & $2$   &  $2$ &  $36^{\circ}$  \tabularnewline
HK Tau   & $2$   &  $5$ &  $32^{\circ}$  \tabularnewline
UZ Tau W & $0.4$ &  $2$ &  $30^{\circ}$  \tabularnewline
\hline
\end{tabular}

\end{center}

\end{table}


\section{Conclusions}

\label{sec:conclusions}

We have presented the line profile models of hydrogens and helium 
based on the axisymmetric MHD simulation of the outflow produced 
at the interface of the magnetosphere of CTTSs and their accretion discs. 
The so-called conical-shell wind (RO09) is formed when the stellar dipole
magnetic field is compressed by the accretion disc into the X-wind
like configuration.  We have extended the previous conical-shell wind
simulations of RO09 to include a more well defined magnetospheric
accretion funnel flow which is essential for modelling the optical and
near-infrared hydrogen and helium lines of CTTSs, as many of the lines
show clear signs of inflow (seen as the inverse P-Cygni profiles).
The formation of the well defined funnel flow with its outer radius
corresponding to $\sim 2\,R_{\odot}$ (Model~B in
Section~\ref{sub:mhd-main-results}) is obtained by lowering the 
disc density or correspondingly the mass-accretion of the disc from
the previously found conical-shell wind solution of RO09 in which the
magnetosphere was compressed all the way to the stellar surface by
the accretion disc. The  mass-accretion and mass-loss rates (in
the conical-shell wind) found in the model (Model~B) are
$4.1\times10^{-8}\,\MsunPerYear$ and $6.9\times10^{-9}\,\MsunPerYear$,
respectively (Table~\ref{tab:MHD-Model-Summary}).  The maximum
velocities in the conical-shell wind and the funnel flows are $\sim
125\,\kmps$ (Fig.~\ref{fig:MHD-Vel}) and $\sim 250\,\kmps$,
respectively. These values are all comparable to the values found in
observations (e.g.~\citealt{hartigan:1995}; \citealt{gullbring:1998},
\citealt{Calvet:2000};  \citealt{alencar:2000}; \citealt{folha:2001};
ED06). In the following, we summarise our main findings from the line
profile models based on this MHD simulation.

We find that \ion{He}{i}~$\lambda10830$ is relatively sensitive to
changes in the X-ray luminosity ($L_{\mathrm{X}}$) and X-ray
temperature ($T_{\mathrm{X}}$) (Section~\ref{sub:depend-X-ray}).  Both
emission and absorption components of the line become stronger as
$L_{\mathrm{X}}$ increases (Fig.~\ref{fig:Lx-test}).  For a fixed
$L_{\mathrm{X}}$ value, the emission and absorption components of the
line become weaker as $T_{\mathrm{X}}$ increases
(Fig.~\ref{fig:Tx-test}) because the peak of the thermal radiation
(assumed as a single temperature gas) shifts to a higher energy and
the amount of the soft X-ray emission decreases. The softer X-ray
photons photoionize more \ion{He}{i} since its photoionization cross
section is larger towards the EUV frequency.  The range of
$L_{\mathrm{X}}$ between $4\times10^{29}$ and
$2\times10^{31}\,\mathrm{erg\,s^{-1}}$ produces reasonable line
strengths in \ion{He}{i}~$\lambda10830$. Similarly, $T_{\mathrm{X}}$
between $2\times 10^6$ and $2\times 10^7\,\Kelvin$ (with
$L_{\mathrm{X}}=4\times10^{30}\,\mathrm{erg\,s^{-1}}$ fixed), also
produced \ion{He}{i}~$\lambda10830$ with its line strength comparable
to observations. On the other hand, yet another wind sensitive line,
H$\alpha$, shows no significant change in its line strength when
$L_{\mathrm{X}}$ and $T_{X}$ are varied in the same ranges used for
\ion{He}{i}~$\lambda10830$ because the photoionization rates (in the
X-ray range for \ion{H}{i}) are small compared to the collisional
rates.

A rich diversity of line profile morphology is found in our sample
hydrogen and helium model profiles (Fig.~\ref{fig:inc-test} in
Section~\ref{sub:Depend-inc}). Many of the model profiles are very
similar to those found in the observations
(e.g.~\citealt{reipurth:1996}; \citealt{edwards:1994};
\citealt{Muzerolle:1998a}; \citealt{alencar:2000};
\citealt{folha:2001}; ED06).    No
wind absorption is seen at a very low inclination angle
(e.g.~$i<22^{\circ}$) in any of the lines considered here.  In
addition, the wind absorption component seen in the higher inclination
angles are rather narrow with the corresponding wind speed
$v \simless 120\,\kmps$. These suggest that the blueshifted absorption
components seen the model profiles are caused by the conical-shell wind and
no significant absorption occurs (at least with the model parameters
used here) in the low-density polar wind.  The main shortcomings of our models are
in the weakness of \ion{He}{i}~$\lambda$5876 emission, and in the
presence of the strong wind absorption component in Pa$\beta$ and
Pa$\gamma$ which are not seen in the observations.

We have examined the relative contributions of the conical-shell wind
and the accretion funnel to the hydrogen and helium line emissions
(Section~\ref{sub:emiss-loc-test}).  We find that the conical-shell
wind can significantly contribute to the line emissions in H$\alpha$,
H$\beta$, Pa$\beta$ and Pa$\gamma$, provided that the temperature in the
wind is high enough (e.g.~$\sim10^{4}\,\Kelvin$). On the other hand,
the helium lines (\ion{He}{i}~$\lambda$10830 and
\ion{He}{i}~$\lambda$5876) have less contribution from the wind
emission. The wind contribution to the hydrogen emission decreases
significantly when a lower wind temperature is adopted in the profile
calculations.  A better constraint of wind temperatures could be
obtained from the comparisons of the line ratios of e.g.~the Balmer
lines from observations and that from line profile models.

Using the various combinations of our main model parameters
($L_{\mathrm{X}}$, $T_{\mathrm{X}}$ and $i$), we have demonstrated
that the conical-shell wind model is capable of producing a subset of 
the line profile morphology for \ion{He}{i}~$\lambda10830$ 
found in observations (Fig.~\ref{fig:compare-obs}). 
The  model is able to reproduce a relatively narrow and low-velocity
blueshifted absorption component in  \ion{He}{i}~$\lambda$10830
profiles, found in about 30~per~cent of the sample in ED09, provided
that the inclination angle is favourable ($\sim 30^{\circ}$ to
$\sim50^{\circ}$ for Model~B, but the range will depend on the opening angle of
the conical-shell wind). However, the model does not reproduce the deep and wide
blueshifted wind absorption components (the P-Cygni like line profile)
found in about 40~per~cent of the sample in ED09.   The earlier
studies by \citet{Kwan:2007} and KU11 demonstrated that the P-Cygni
like profiles seen in  \ion{He}{i}~$\lambda$10830 of CTTSs are most
likely produced by the absorption by the stellar wind that blows in
the polar directions. The inability of our model for reproducing
the P-Cygni like profile was expected since our MHD model did not
include the stellar wind component.

Finally, in a future study, we plan to implement the stellar wind in
our MHD simulations, and study its effect on the line formations,
especially on the P-Cygni like profile of \ion{He}{i}~$\lambda$10830. 
We also plan to investigate the line formation problems in the higher
mass-accretion regimes which are somewhat similar to Model~A
(Fig.~\ref{fig:MHD01} in Section~\ref{sec:mhd-results}), and those in
\citet{Koenigl:2011} and \citet{Lii:2012}. 
These models, in which the compression of the magnetosphere toward the
stellar surface is strong and the outflow due to the magnetic
pressure gradient force is very strong,  may be applicable to the
objects during the outbursts periods such as EXors and FUORs. 
The time-series line profiles of the outburst can be modelled by using 
multiple time slices (rather than a single time slice used in this
work) of MHD simulations, to study the time evolution of observed
spectra.

\section*{Acknowledgements} 

We thank an anonymous referee who provided us valuable comments and
suggestions which helped improving the manuscript. We thank Tim
Harries for providing us valuable comments on the 
manuscript.  We are grateful for Susan Edwards for providing us the
observations of \ion{He}{i}~$\lambda10830$ and Pa$\gamma$ line
profiles. RK would like to thank University of Nevada, Las Vegas for
their hospitality during his visit. 
The research conducted by RK and MMR is supported by NASA
grant NNX10AF63G, NNX11AF33G and NSF grant AST-1008636. Resources
supporting this work were provided by the NASA High-End Computing
(HEC) Program through the NASA Advanced Supercomputing (NAS) Division
at Ames Research Center and the NASA Center for Computational
Sciences (NCCS) at Goddard Space Flight Center.


\end{document}